\documentclass[longauth]{aa} 
\usepackage{graphicx}
\usepackage{tabularx}
\usepackage{multirow}
\usepackage[colorlinks=true, linkcolor=blue, citecolor=blue, draft=False]{hyperref}
\usepackage{gensymb}
\usepackage{color}
\usepackage{txfonts}
\usepackage{natbib}
\usepackage{amsmath}
\usepackage{ulem}
\usepackage[flushleft]{threeparttable}
\usepackage[export]{adjustbox}

\defcitealias{Wang2018}{W18}
\defcitealias{Wang2016}{W16}

\begin{document}

\title{Starbursts with suppressed velocity dispersion revealed in a forming cluster at $\MakeLowercase{z}=2.51$}
\author{M.-Y.~Xiao\inst{\ref{inst1},\ref{inst2},\ref{inst3}}
\and T. Wang\inst{\ref{inst1},\ref{inst3}}\thanks{E-mail: taowang@nju.edu.cn; mengyuan.xiao@cea.fr}
\and D. Elbaz\inst{\ref{inst2}} 
\and D. Iono\inst{\ref{inst4},\ref{inst5}} 
\and X. Lu\inst{\ref{inst4}} 
\and L.-J. Bing\inst{\ref{inst6}} 
\and E. Daddi\inst{\ref{inst2}} 
\and B. Magnelli\inst{\ref{inst2}} 
\and C. G\'omez-Guijarro\inst{\ref{inst2}} 
\and F. Bournaud\inst{\ref{inst2}} 
\and Q.-S. Gu\inst{\ref{inst1},\ref{inst3}}
\and S. Jin\inst{\ref{inst7},\ref{inst8}}
\and F. Valentino\inst{\ref{inst7},\ref{inst9}}
\and A. Zanella\inst{\ref{inst10}}  
\and R. Gobat\inst{\ref{inst11}} 
\and S. Martin\inst{\ref{inst12},\ref{inst13}}
\and G. Brammer\inst{\ref{inst7},\ref{inst9}}
\and K. Kohno\inst{\ref{inst14}}  
\and C. Schreiber\inst{\ref{inst15}} 
\and L. Ciesla\inst{\ref{inst6}} 
\and X.-L. Yu\inst{\ref{inst1},\ref{inst3}} 
\and K. Okumura\inst{\ref{inst2}} 
}

\institute{School of Astronomy and Space Science, Nanjing University, Nanjing 210093,P. R. China\label{inst1}
\and AIM, CEA, CNRS, Universit\'e Paris-Saclay, Universit\'e Paris Diderot, Sorbonne Paris Cit\'e, F-91191 Gif-sur-Yvette, France\label{inst2}
\and Key Laboratory of Modern Astronomy and Astrophysics (Nanjing University), Ministry of Education, Nanjing 210093, China\label{inst3}
\and National Astronomical Observatory of Japan, 2-21-1 Osawa, Mitaka, Tokyo 181-8588, Japan\label{inst4}
\and Department of Astronomical Science, SOKENDAI (The Graduate University for Advanced Studies), Mitaka, Tokyo 181-8588, Japan\label{inst5}
\and Aix Marseille Universit\'e, CNRS, LAM, Laboratoire d’Astrophysique de Marseille, Marseille, France\label{inst6}
\and Cosmic Dawn Center (DAWN)\label{inst7}
\and DTU-Space, Technical University of Denmark, Elektrovej 327, DK-2800 Kgs. Lyngby, Denmark\label{inst8}
\and Niels Bohr Institute, University of Copenhagen, Blegdamsvej 17, DK2100 Copenhagen {\O}, Denmark\label{inst9}
\and Istituto Nazionale di Astrofisica, Vicolo dell’Osservatorio 5, I-35122 Padova, Italy\label{inst10}
\and Instituto de F\'isica, Pontificia Universidad Cat\'olica de Valpara\'iso, Casilla 4059, Valpara\'iso, Chile\label{inst11}
\and European Southern Observatory, Alonso de C\'ordova, 3107, Vitacura, Santiago 763-0355, Chile\label{inst12}
\and Joint ALMA Observatory, Alonso de C\'ordova, 3107, Vitacura, Santiago 763-0355, Chile\label{inst13}
\and Institute of Astronomy, Graduate School of Science, The University of Tokyo, 2-21-1 Osawa, Mitaka, Tokyo 181-0015, Japan\label{inst14}
\and Astrophysics, Department of Physics, Keble Road, Oxford OX1 3RH, UK\label{inst15}
}

\abstract
{One of the most prominent features of galaxy clusters is the presence of a dominant population of massive ellipticals in their cores. 
Stellar archaeology suggests that these gigantic beasts assembled most of their stars in the early Universe via starbursts.  
However, the role of dense environments and their detailed physical mechanisms in triggering starburst activities remain unknown. Here we report spatially resolved Atacama Large Millimeter/submillimeter Array (ALMA) observations of the CO $J= 3-2$ emission line, with a resolution of about 2.5 kiloparsecs, toward a forming galaxy cluster core with starburst galaxies at $z=2.51$.  In contrast to starburst galaxies in the field often associated with galaxy mergers or highly turbulent gaseous disks, our observations show that the two starbursts in the cluster exhibit dynamically cold (rotation-dominated) gas-rich disks. Their gas disks have extremely low velocity dispersion ($\sigma_{\mathrm{0}} \sim 20-30$ km s$^{-1}$), which is three times lower than their field counterparts at similar redshifts.  The high gas fraction and suppressed velocity dispersion yield gravitationally unstable gas disks, which enables highly efficient star formation. The suppressed velocity dispersion, likely induced by the accretion of corotating and coplanar cold gas, might serve as an essential avenue to trigger starbursts in massive halos at high redshifts.}

   \maketitle

\section{Introduction}

Galaxy clusters represent the densest environments and trace the most massive dark matter halos in the Universe \citep{Kravtsov2012}. It is well known that massive elliptical galaxies are enhanced in clusters \cite[e.g.,][]{Dressler1980,Peng2010}. However, it remains unclear whether and how the dense environment affects their formation and quenching. To answer this question, we need to trace them back to the cosmic epoch when they are still actively forming stars. 

During the last decade, a number of young clusters and protoclusters have been found at $z > 2$ \citep[e.g.,][]{Wang2016, Casey2016, Miller2018, Oteo2018}, the peak of cosmic star formation history \citep[][]{Madau2014}. Contrary to mature clusters in the local Universe, which are dominated by massive quiescent galaxies in their cores \citep[e.g.,][]{Dressler1980}, these young clusters and protoclusters host a significant population of massive star-forming galaxies. In particular, observations have revealed the ubiquity of starbursts in these high-$z$ (proto)clusters. For example, \citet{Miller2018} observed a protocluster at $z=4.3$ hosting at least 14 gas-rich galaxies in a projected region of 130 kiloparsecs (kpc) in diameter with a total star formation rate (SFR) of 6000 M$_\odot$ yr$^{-1}$; \cite{Wang2016} identified an X-ray cluster at $z=2.51$ with a SFR of about 3400 M$_\odot$ yr$^{-1}$ in the central 80 kpc region.  Many observations have found starbursting overdensities in the early Universe \citep[e.g.,][]{Blain2004,Casey2015,Casey2016,Oteo2018}, which are in line with the expectation of the hierarchical growth of structures associated with an enhancement of star formation or starburst galaxies in dense environments  \citep{Dannerbauer2014,Dannerbauer2017,Hayashi2016}.

However, the triggering mechanism of high-$z$ starbursts in dense environments is unclear. Observations have revealed evidence of merger and/or interactions in these starbursts in (proto)clusters  \citep[e.g.,][]{Coogan2018,Hodge2013}, 
while simulations show the inefficiency of enhancing star formation in high-$z$ galaxies, even in the event of the most drastic gas-rich major mergers  \citep{Fensch2017}.
Meanwhile, starburst galaxies in dense environments such as GN20, similar to those in the field, contain a rotating gas disk and do not exhibit major merger evidence \citep{Hodge2012}.

To understand the origin of starbursts in dense environments in the high-$z$ Universe, spatially and spectroscopically resolved measurements of their gas kinematic are needed. However, the large cost of high-resolution and high-sensitivity observations of galaxies in the early Universe has always been a limitation for in-depth kinematic studies of high-$z$ starbursts in (proto)clusters. 

\begin{figure*}[h!]
        \begin{center}
        \includegraphics[width=1\textwidth]{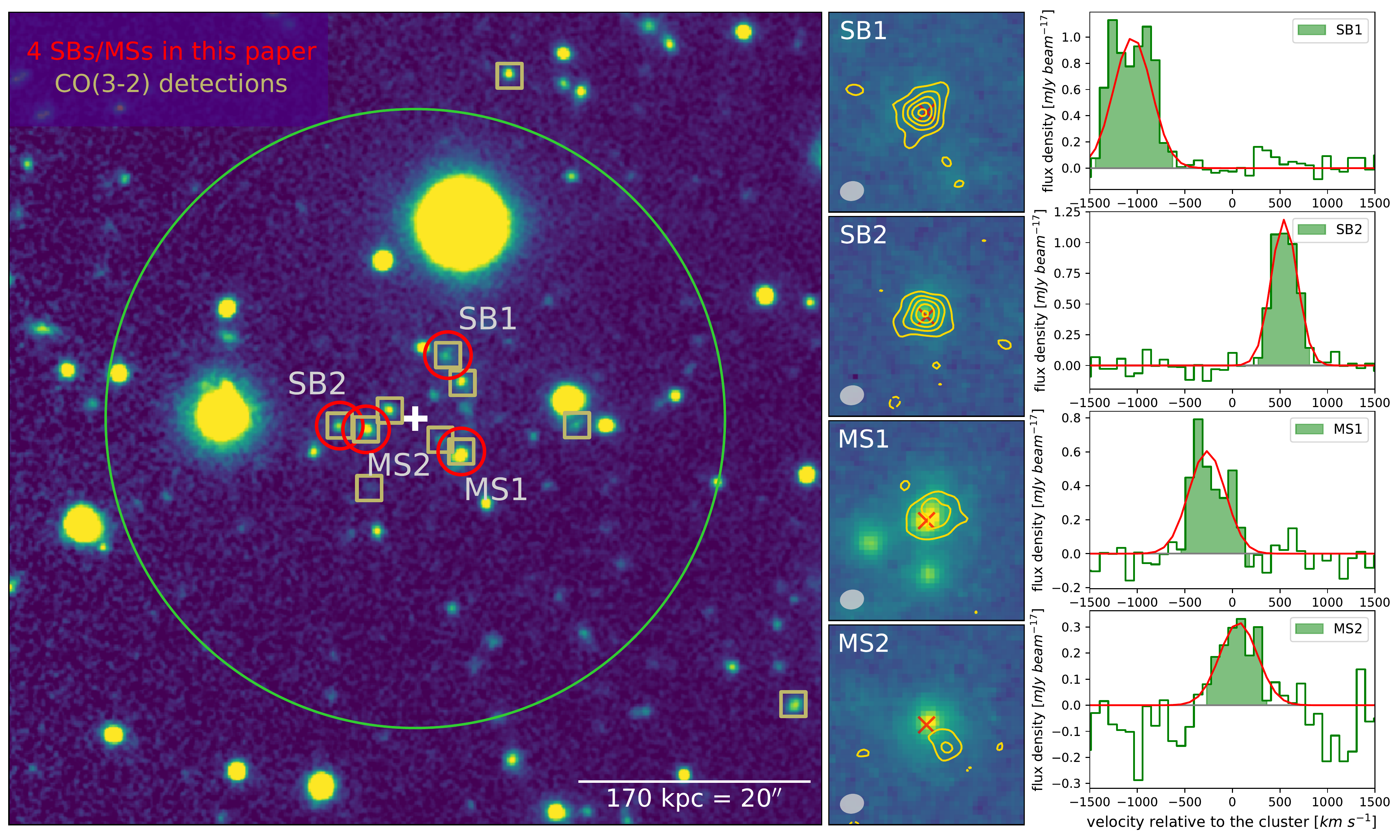}
         \end{center}
          \vspace{-0.6truecm}  
        \caption{Galaxy cluster CLJ1001 at $z=2.51$ and the four member galaxies.
\textbf{Left}: Sky distributions of member galaxies around the cluster center. Red circles mark the two SBs and two MSs with the brightest CO(3$-$2) luminosities ($S/N\gg10$; $S/N\sim30$ for the two SBs) among all CO(3$-$2) detections (golden squares) in the central region of the cluster.   The background is the $K_{\rm s}$-band image from the UltraVista survey  
with a size of 70$^{\prime\prime}$ $\times$ 70$^{\prime\prime}$. The white cross shows the cluster center. The scale bar indicates  half of the virial radius ($R_{\rm 200c}$ $\sim 340$ kpc) of the cluster. The large green circle denotes the ALMA FOV, corresponding to FWHP $\sim$ 53$^{\prime\prime}$ of the ALMA antennas' primary beam at 98.63 GHz. 
\textbf{Middle}: Velocity-integrated intensity map ($Moment~0$) of CO(3$-$2) (golden contours) detected by ALMA overlaid on the \textit{HST}/F160W image of the two SBs and two MSs. Each panel is 2.5$^{\prime\prime}$ $\times$ 2.5$^{\prime\prime}$. The angular resolution is $0.31^{\prime\prime} \times 0.25^{\prime\prime}$ (gray-filled ellipse in the bottom-left corner). The contour levels start at $\pm3\sigma$  and increase in steps of $\pm3 \sigma$, where positive and negative contours are solid and dashed, respectively. The red cross in each panel denotes the centroid of the stellar emission determined from the \textit{HST}/F160W image. The derived integrated fluxes are presented in Table \ref{tab1}.
\textbf{Right}: CO(3$-$2) line spectra of the four member galaxies. The CO lines are binned at 90 km s$^{-1}$, and the velocity range is shaded in green over which $Moment~0$ maps are integrated. The best-fit single Gaussian profiles are overlaid in red.}
        \label{fig1}
        \vspace{-12pt}
\end{figure*}

One of the most distant young clusters, CLJ1001 at $z_{\rm spec}=2.51$ (\citealp{Wang2016}, W16, hereafter; see also \citealp{Casey2015,Daddi2017,Wang2018,Cucciati2018,Gomez-Guijarro2019,Champagne2021}), exhibits extended X-ray emission and encompasses an overdensity of massive star-forming galaxies, with a total SFR of $\sim$ 3400 $M_{\odot}$yr$^{-1}$ in its 80 kpc core (the cluster virial radius is $R_{\rm 200c} \sim 340$ kpc) and a gas depletion time of $\sim$ 200 Myr. CLJ1001 is in the phase of rapid transformation from a protocluster to a mature cluster, which makes it an ideal laboratory to study the connection between the triggering mechanism of cluster starbursts and their dense environment.  
Revealing the physical origin of cluster starbursts is key to uncovering the formation and quenching mechanisms of massive galaxies in clusters, which has been a longstanding problem in extragalactic studies. With these aims, we have conducted high-resolution CO $J= 3-2$ emission line (CO(3$-$2), hereafter) observations with the Atacama Large Millimeter/submillimeter Array (ALMA) toward the cluster core of CLJ1001 (Fig. \ref{fig1}). Thanks to newly obtained ALMA high resolution data, in this paper we study the molecular gas spatial distribution and resolved kinematics of two massive starburst galaxies (SBs) and two star-forming main sequence galaxies (MSs) in the galaxy cluster CLJ1001.

This paper is organized as follows: in $\S$\ref{Sec:data}, we introduce the ALMA CO(3$-$2) line and 3.2 mm continuum observations. In $\S$\ref{Sec:Data Analysis}, we study the two cluster SBs and two cluster MSs involved in this work and analyze the structural properties of their molecular gas, CO excitation, dust mass, molecular gas mass, and gas kinematics.  In $\S$\ref{Sec:results}, we present the results focusing on their molecular gas kinematics and their gas disk stabilities. In $\S$\ref{Discussion}, we discuss the triggering mechanism of the cluster SBs and the drivers of the low turbulent gas in the two cluster SBs. We summarize the main conclusions in $\S$\ref{Sec:summary}.

We adopt a  \cite{Chabrier2003} initial mass function (IMF) to estimate the SFR and stellar mass. We assume cosmological parameters of $H_{0}$ = 70 km s$^{-1}$ Mpc$^{-1}$, $\Omega_{M}$ = 0.3, and $\Omega_{\Lambda}$ = 0.7. When necessary, data from the literature have been converted with  a conversion factor 
of SFR \citep[][IMF]{Salpeter1955} = 1.7  $\times$ SFR  \citep[][IMF]{Chabrier2003} and M$_*$ \citep[][IMF]{Salpeter1955} = 1.7  $\times$ M$_*$ \citep[][IMF]{Chabrier2003}.

\section{ALMA observations}
\label{Sec:data}

We carried out observations of the CO(3$-$2) transitions at the rest-frame frequency of $345.796 \,{\rm GHz}$ ($98.630 \,{\rm GHz}$ in the observed frame) for the galaxy cluster CLJ1001 at $z=2.51$ with ALMA band-3 receivers. These ALMA Cycle 4 observations (Project ID: 2016.1.01155.S; PI: T.~Wang) were taken between 2016 November and 2017 August with a total observing time, including calibration and overheads, of approximately 7 hours. 
We adopted two different array configurations to obtain reliable measurements of total flux and resolved kinematics information: 
 a more compact array configuration (C40-4) observing low-resolution large spatial scales, while a more extended array configuration (C40-7) observing high-resolution small spatial scales, with an on-source time of 1.6 hours and 2.2 hours, respectively. 
 The observations were performed with a single pointing covering a field of view (FOV) of $\sim$ 53$^{\prime\prime}$, corresponding to the full-width at half power (FWHP) of the ALMA primary beam.

The calibration was performed using version 5.3.0 of the Common Astronomy Software Application package (CASA) \citep{mcmullin} with a standard pipeline. We carried out the data reduction in two cases. Case 1 was designed to obtain a high-resolution cube of CO(3$-$2) line. To this end, we first combined two configurations' data to form a single visibility table (UV table), using visibility weights of the compact configuration dataset (C40-4) to the extended configuration dataset (C40-7)  proportional to 1:4. Imaging was carried out using the $tclean$ task with 0.04$^{\prime\prime}$ pixels and a channel width of 30 km s$^{-1}$ with a Briggs weighting of robust = 0.5 scheme. The resulting data cube has a synthesized beam size of  $0.31^{\prime\prime} \times 0.25^{\prime\prime}$ ($\sim2.5$ kpc $\times$ 2.0 kpc in physical scale) with an rms sensitivity of $\sim 110-120~\mu$Jy beam$^{-1}$ per channel at the phase center. Case 2 was designed to derive the total flux of our sources. To this end, we combined two configurations' data with visibility weights proportional to 1:1. We used 0.2$^{\prime\prime}$ pixels and a channel width of 90 km s$^{-1}$ with a uvtaper of 1$^{\prime\prime}$ and a natural weighting scheme for imaging. The natural weighting provides a better sensitivity, enabling us to measure the total flux more robustly.
The resulting data cube has a synthesized beam size of $1.54^{\prime\prime} \times 1.37^{\prime\prime}$ ($\sim12.3$ kpc $\times$ 11.0 kpc in physical scale) and a central rms level of $\sim 60~\mu$Jy beam$^{-1}$ per channel.

We also created the observed 3.2 mm continuum maps using Briggs weighting (robust $=$ 0.5) after excluding the frequency range of the CO(3$-$2) line. We used the same method as for the CO(3$-$2)  cubes to create two continuum maps with different angular resolutions. To estimate the continuum fluxes of our sources, we created continuum maps with a combination of two configurations' data with visibility weights proportional to 1:1. The rms level is $\sim 4.3~\mu$Jy beam$^{-1}$ in the map of $1.12^{\prime\prime} \times 1.06^{\prime\prime}$ angular resolution ($\sim9.0$ kpc $\times$ 8.5 kpc in physical scale). To measure the dust continuum sizes of our sources, we created continuum maps with a combination of two configurations' data with visibility weights of the compact configuration dataset (C40-4) to the extended configuration dataset (C40-7) proportional to 1:4. The rms level is $\sim 6.9~\mu$Jy beam$^{-1}$ in the map of $0.30^{\prime\prime} \times 0.24^{\prime\prime}$ angular resolution ($\sim2.4$ kpc $\times$ 1.9 kpc in physical scale).

  \begin{table*}\footnotesize     
\caption{Physical properties of the two SBs and two MSs in CLJ1001.}             
\centering
  
  \begin{tabular}{c cccccc}
    \hline
    \hline
    \noalign{\medskip}
     ID&  $^a$ID$_{K_{\rm s}}$ &R.A.&Decl.  &  $z_{\rm CO(3-2)}$ & $I_{\rm CO(3-2)}$ & FWHM$_{\rm CO(3-2)}$     \\
      && (J2000)&(J2000)& & (Jy km s$^{-1}$) &  (km s$^{-1}$)    \\
        \noalign{\medskip}
        \hline
        \noalign{\smallskip}

  SB1 & 131077&10:00:56.95& +02:20:17.2& 2.494  &1.273 $\pm$ 0.036& 547 $\pm$ 38   \\
  SB2  & 130891&10:00:57.56&+02:20:11.2& 2.512  &0.764 $\pm$ 0.028&  324 $\pm$ 17    \\
  MS1  &130949&10:00:56.86&+02:20:08.7& 2.503  &0.537 $\pm$ 0.023& 453 $\pm$ 47  \\
 MS2  &130901&10:00:57.39&+02:20:10.8& 2.507  &0.242 $\pm$ 0.024&  472 $\pm$ 103  \\

 \hline
\\
\hline
\hline
$^{b}$ log $M_{\rm *}$ & $^{c}$log $L_{\rm IR}$ &$^{d}$SFR& $^{e} L^{\prime}_{\rm CO(1-0)}$& $L^{\prime}_{\rm CO(3-2)}$&$^{f} R_{\rm 31}$&$S_{\rm 3.2mm}$ \\
($M_{\odot}$)& ($L_{\odot}$)&($M_{\odot}$yr$^{-1}$) & (10$^{10}$ K km s$^{-1}$ pc$^{2}$) & (10$^{10}$ K km s$^{-1}$ pc$^{2}$) && ($\mu$Jy) \\
  \hline
10.93 $\pm$ 0.15& 12.95$^{+0.04}_{-0.04}$ & 1314 $\pm$ 122 & 4.9 $\pm$ 0.4&  4.10 $\pm$ 0.12& 0.84 $\pm$ 0.07 &82 $\pm$ 10 \\ 
10.83 $\pm$ 0.15 & 12.70$^{+0.16}_{-0.26}$ & 751 $\pm$ 338  &3.2 $\pm$ 0.3&  2.49 $\pm$ 0.09& 0.78 $\pm$ 0.08&38 $\pm$ 10  \\
11.36 $\pm$ 0.15& 12.29$^{+0.16}_{-0.25}$ & 292 $\pm$ 128 & 2.3 $\pm$ 0.2& 1.74 $\pm$ 0.07 & 0.76 $\pm$ 0.07&62 $\pm$ 12 \\
 11.35 $\pm$ 0.15&12.30$^{+0.20}_{-0.40}$ & 300 $\pm$ 179 & 1.8 $\pm$ 0.3& 0.79 $\pm$ 0.08& 0.44 $\pm$ 0.08&18 $\pm$ 3  \\

\hline
\\
\hline
\hline
\multicolumn{1}{c} {$^{g}M_{\rm dust}$}&\multicolumn{2}{c} {$^{h}$CO(3$-$2) Size ($UV$ $plane$)}&\multicolumn{2}{c} {$^{i}$CO(3$-$2) size ($image$ $plane$)} &\multicolumn{2}{c} {$^{j}$3.2 mm size (image plane)}\\
\multicolumn{1}{c} {($10^{8}$$M_{\odot}$)}&\multicolumn{2}{c} {(kpc $\times$ kpc)}& \multicolumn{2}{c} {(kpc $\times$ kpc)} & \multicolumn{2}{c} {(kpc $\times$ kpc)}  \\
        \noalign{\medskip}
        \hline
        \noalign{\smallskip}
\multicolumn{1}{c} {16.0 $\pm$ 1.5}&\multicolumn{2}{c} {(2.05 $\pm$ 0.10) $\times$ (0.98 $\pm$ 0.11)}&\multicolumn{2}{c} {(2.10 $\pm$ 0.31) $\times$ (1.43 $\pm$ 0.32)}&\multicolumn{2}{c} {(1.87 $\pm$ 0.54)$\times$(1.27 $\pm$ 0.54)} \\
\multicolumn{1}{c} {10.8 $\pm$ 2.2}&\multicolumn{2}{c} {(1.42 $\pm$ 0.11) $\times$ (1.14 $\pm$ 0.17)}&\multicolumn{2}{c} {(1.70 $\pm$ 0.30) $\times$ (1.40 $\pm$ 0.32)}& \multicolumn{2}{c} {(...)}\\
\multicolumn{1}{c} {8.3 $\pm$ 4.8}&\multicolumn{2}{c} {(1.77 $\pm$ 0.20) $\times$ (1.56 $\pm$ 0.33)}&\multicolumn{2}{c} {(2.13 $\pm$ 0.32) $\times$ (1.79 $\pm$ 0.29)}&\multicolumn{2}{c} {(...)}\\
\multicolumn{1}{c} {4.3 $\pm$ 1.0}&\multicolumn{2}{c} {(0.79 $\pm$ 0.20) $\times$ (0.79 $\pm$ 0.37)}&\multicolumn{2}{c} {(2.32 $\pm$ 0.60) $\times$ (0.90 $\pm$ 0.53)}&\multicolumn{2}{c} {(...)}\\

    \hline
    \end{tabular}
\tablefoot{$^{a}$IDs from the $K_{\rm s}$-selected catalog \citep{Muzzin2013}. $^{b}$$M_{\rm *}$, derived from the SED fitting in \citetalias{Wang2016} (scaled to a \citealt{Chabrier2003} IMF by a factor 1.7). $^{c}$Total infrared luminosity, derived from the infrared SED fitting with \texttt{CIGALE}. $^{d}$SFR, derived from $L_{\rm IR}$ using \cite{kennicutt} (${\rm SFR}=\,1.49 \times 10^{-10} L_{\rm IR}$; \citealt{Chabrier2003} IMF). $^{e}L^{\prime}_{\rm CO(1-0)}$, from \citetalias{Wang2018}. $^{f}$CO excitation: $R_{\rm 31} = L^{\prime}_{\rm CO(3-2)}/L^{\prime}_{\rm CO(1-0)}$. $^{g}M_{\rm dust}$, derived from the infrared SED fitting with \texttt{CIGALE}. $^{h}$The best-fit semi-major and semi-minor axes of CO(3$-$2) in the $UV$ plane. $^{i}$The deconvolved semi-major and semi-minor axes of CO(3$-$2) in the image plane. $^{j}$The deconvolved semi-major and semi-minor axes of 3.2 mm dust continuum emission in the image plane.}
\label{tab1}
\end{table*}

\begin{table*}\footnotesize     
\caption{Gas properties of the two SBs and two MSs.}
\centering
\begin{tabular}{ c c c c c} 
\hline\hline
                \noalign{\smallskip}
                \multicolumn{1}{c} {} & SB1 & SB2 & MS1 & MS2\\
                \noalign{\medskip}
                \hline
                \noalign{\smallskip}
                $^{a} \alpha_{\rm CO}(Z)$ ($M_{\odot}$/(K km s$^{-1}$ pc$^{2}$))   & 4.09 & 4.10 & 4.06 & 4.06 \\
                $^{b} M_{\rm gas,CO}$ ($10^{10}$$M_{\odot}$) & 20.2 $\pm$ 1.7& 13.3 $\pm$ 1.4 & 9.2 $\pm$ 0.9 & 7.4 $\pm$ 1.1\\
                 $^{c} f_{\rm gas,CO}$ & 0.70$^{+0.09}_{-0.06}$ & 0.66$^{+0.10}_{-0.07}$ & 0.29$^{+0.09}_{-0.06}$ & 0.25$^{+0.08}_{-0.06}$\\
                  $^{d} t_{\rm dep,CO}$ (Gyr)& 0.15 $\pm$ 0.02 & 0.18 $\pm$ 0.08 & 0.32 $\pm$ 0.14 & 0.25 $\pm$ 0.15 \\
                  \\
                  $^{e}\delta_{\rm GDR}$ & 126 $\pm$ 16 & 123 $\pm$ 28 & 110 $\pm$ 65 & 171 $\pm$ 47\\
                  $^{f} M_{\rm gas,GDR}$ ($10^{10}$$M_{\odot}$) & 17.9 $\pm$ 6.4 & 12.8 $\pm$ 5.1 & 7.6 $\pm$ 5.1 & 4.0 $\pm$ 1.7\\
                  $ f_{\rm gas,GDR}$ & 0.68$^{+0.12}_{-0.10}$ & 0.65$^{+0.13}_{-0.11}$ & 0.25$^{+0.15}_{-0.14}$ & 0.15$^{+0.08}_{-0.07}$\\
                 $ t_{\rm dep,GDR}$ (Gyr) &  0.14 $\pm$ 0.05&  0.17 $\pm$ 0.10&  0.26 $\pm$ 0.21&  0.13 $\pm$ 0.10\\
                 \\
                  $^{g} M_{\rm gas,3.2mm}$ ($10^{10}$$M_{\odot}$) & 22.7 $\pm$ 6.4 & 10.4 $\pm$ 3.9 & 17.1 $\pm$ 5.5 & 5.1 $\pm$ 1.6\\
                  $ f_{\rm gas,3.2mm}$ & 0.73$^{+0.10}_{-0.08}$ & 0.61$^{+0.13}_{-0.11}$ & 0.43$^{+0.13}_{-0.11}$ & 0.18$^{+0.08}_{-0.07}$\\
                 $ t_{\rm dep,3.2mm}$ (Gyr) &  0.17 $\pm$ 0.05&  0.14 $\pm$ 0.08&  0.59 $\pm$ 0.32&  0.17 $\pm$ 0.11\\
\hline
\end{tabular}

\tablefoot{The molecular gas masses are estimated based on the CO(1$-$0) emission line using metallicity-dependent conversion factors, gas-to-dust ratio, and 3.2mm dust continuum emission. $^{a}$CO-to-H$_2$ conversion factor ($\alpha_{\rm CO}(Z)$) from \citetalias{Wang2018}, calculated based on the mass-metallicity relation. $^{b}$Total molecular gas mass, computed as $M_{\rm gas,CO} = \alpha_{\rm CO}(Z) L_{\rm CO(1-0)}^{\prime}$. $^{c}$Gas fraction: $f_{\rm gas} = M_{\mathrm{gas}}/(M_{\mathrm{star}} + M_{\mathrm{gas}})$. $^{d}$Gas depletion time: $t_{\rm dep} = M_{\mathrm{gas}}/SFR$, which is the inverse of the star formation efficiency (SFE $= 1/t_{\rm dep}$). $^{e}$Gas-to-dust mass ratio, computed based on the $\delta_{\rm GDR}-Z$ relation (Eq.~\ref{GDR}). $^{f}$$M_{\rm gas,GDR}$, computed based on the gas-to-dust ratio (Eq.~\ref{Mgas}). $^{g}$$M_{\rm gas,3.2mm}$, computed based on the Rayleigh-Jeans tail dust continuum (Eq.~\ref{RJ}).}
\label{tab3}
\end{table*}

\section{Data analysis}
\label{Sec:Data Analysis}

\subsection{Starburst and main-sequence galaxies}

Four member galaxies in the cluster CLJ1001 have a CO(3$-$2) line detected with high significances ($35\sigma$, $27\sigma$, $23\sigma$, and $10\sigma$; see Table \ref{tab1}), allowing a detailed study of their kinematics. They are all massive star-forming galaxies with stellar masses of $\log(M_*/M_\odot)>10.8$, and they are located in the central 80 kpc region of the cluster (Fig. \ref{fig1}).
 These include two SBs and two MSs, with the SBs exhibiting specific star-formation rates (sSFR $\equiv$ SFR$/M_*$) more than three times the star-forming main-sequence \citep[SFMS;][]{Schreiber2015}.  According to their distances from the SFMS ($\Delta$MS = SFR/SFR$_{\rm MS}$ $\sim$ 5.8, 3.9, 0.7, and 0.7, where SFR$_{\rm MS}$ is the SFR of a galaxy on the MS with the same stellar mass), the four galaxies were then named SB1, SB2, MS1, and MS2, respectively. The stellar masses were derived from spectral energy distribution (SED) fitting \citep[\citetalias{Wang2016};][W18, hereafter]{Wang2018}. The SFRs were calculated from the infrared luminosity \citep{kennicutt}\footnote{${\rm SFR}=\,1.49 \times 10^{-10} L_{\rm IR}$ in \cite{kennicutt}, which uses a \citealt{Kroupa2001} IMF. Here we neglect the small differences between the \citealt{Kroupa2001} IMF and the \citealt{Chabrier2003} IMF, and assume the same SFRs derived based on the two IMFs.}, which was derived from our infrared SED fitting (updated version of \citetalias{Wang2016} and \citetalias{Wang2018} with the addition of 3.2mm data, see $\S$\ref{dustmass}).  The resulting SFRs of these four galaxies are consistent with those derived from the infrared luminosity of \citetalias{Wang2016} and \citetalias{Wang2018} within a 1$\sigma$ confidence level. We note that all data presented here have been converted to a \citealt{Chabrier2003} IMF. The results are presented in  Table \ref{tab1}.

The intense star formation of the two SBs shows that they are rapidly building up their stellar masses at a rate of 1314 $\pm$ 122 $M_{\odot}$yr$^{-1}$ and  751 $\pm$ 338 $M_{\odot}$yr$^{-1}$ (see Table \ref{tab1}).  The CO(3$-$2) emission of all four galaxies reveals continuous velocity gradients in the observed gas rotation velocity fields ($Moment~1$) and position-velocity (PV) diagrams (Fig. \ref{fig2}). The observed velocity dispersion fields ($Moment~2$) exhibit central dispersion peaks. These are consistent with the kinematics of rotating disks.

\subsection{Structural properties of the molecular gas}
We measured the molecular gas structural properties of the four cluster members. To avoid uncertainties from the imaging process, we derived the source sizes with the visibility CO(3$-$2) data in the $UV$ plane by fitting an elliptical Gaussian (task $uvmodelfit$). The best-fit semi-major and semi-minor axes for the SB1, SB2, MS1, and MS2 are shown in Table \ref{tab1}.  As a comparison, we also performed a two-dimensional (2D) elliptical Gaussian fit on the high-resolution $Moment~0$ map of CO(3$-$2). The deconvolved semi-major and semi-minor axis values were broadly consistent with our $UV$ analysis. We note that MS2 shows different sizes in the imaging and $UV$ analyses, which could be caused by its lowest significance ($10\sigma$) of (3$-$2) detections among the four sources. The two SBs with higher significances of CO(3$-$2) detections than the two MSs show intrinsic sizes consistent in both methods within the error. 

For the dust size, we only successfully measured the SB1 in the image plane, which has the highest signal-to-noise ratio ($S/N\sim8$) of 3.2 mm dust continuum emission among the four sources. The dust size of the SB1 was measured by a 2D elliptical Gaussian fit on the high-resolution $Moment~0$ map at 3.2 mm. We found that the dust size and molecular gas size of the SB1 are consistent within the error (see Table \ref{tab1}). Assuming a lack of significant size variation between the CO and dust continuum \citep[e.g.,][]{Puglisi2019}, we compared the molecular gas size (and/or dust size) of our sources with the literature. The typical dust size for galaxies at $z=2.5$ with a stellar mass of $10^{10.9} M_\odot$ is $0.99\pm0.34$ kpc in radius from the GOODS-ALMA 1.1 mm survey \citep[Figure 15 in][]{Gomez-Guijarro2021}. The dust distribution in the GOODS-ALMA sample was considered to be compact relative to the stellar distribution \citep{vanderWel2014} in \cite{Gomez-Guijarro2021}. By further using the submillimeter compactness criterion \citep{Puglisi2021}, which is used to select compact galaxies, defined as a ratio of the stellar size \citep{vanderWel2014} larger than the molecular gas size by a factor of 2.2, we conclude that the two SBs do not show compact gas disks.

\subsection{CO excitation}
The CO(3$-$2) line flux was measured from a 2D Gaussian fitting on the velocity-integrated intensity map ($Moment~0$) with the velocity range shown in Fig. \ref{fig1}. Then we calculated its luminosity $L_{\mathrm{CO(3-2)}}^{\prime}$  \citep{Solomon2005}. The CO(1$-$0) line luminosity $L_{\rm CO(1-0)}^{\prime}$ has already been obtained with the Karl G. Jansky Very Large Array (JVLA) observations in \citetalias{Wang2018}. Therefore, we can derive the CO excitation $R_{\rm 31}$ with the CO(3$-$2) to CO(1$-$0) line luminosity ratio ($R_{\rm 31}  = L_{\rm CO(3-2)}^{\prime}/L_{\rm CO(1-0)}^{\prime}$). The values are listed in Table \ref{tab1}. The $R_{\rm 31}$ of the two SBs is around 0.8. 

It has been shown that starburst galaxies generally have higher CO excitation rates relative to main-sequence objects \citep[e.g.,][]{Valentino2020,Puglisi2021}. However, there exists a large variation in the $R_{\rm 31}$ values of high-$z$ galaxies across the literature. In \cite{Riechers2020}, the median $R_{\rm 31}$ is $0.84 \pm 0.26$ for MS galaxies at $z=2-3$ in the VLASPECS survey, while \cite{Daddi2015} showed an $R_{\rm 31}$ of $0.42 \pm 0.07$ for MS galaxies at $z\sim1.5$, which is a factor of two smaller. In addition, \cite{Sharon2016} presented an $R_{\rm 31}$ of $0.78 \pm 0.27$ for dusty submillimeter galaxies at $z \sim$ 2, while \cite{Harrington2018} reported an $R_{\rm 31}$ of only 0.34 for the strongly lensed hyper-luminous infrared galaxies at $z \sim$ 1-3. 
 For starburst galaxies in protoclusters at similar redshifts, they also exhibit a broad range of $R_{\rm 31}\sim0.5-1.2$ (e.g., HXMM20 at $z=2.6$ in \citealp{Gomez-Guijarro2019} and BOSS 1441  at $z=2.3$ in \citealp{Li2021}). Therefore, 
 the measured $R_{\rm 31}$ values for the two SBs are consistent with a wide range of values observed in other galaxies (including MSs and starbursts) at similar redshifts \citep[e.g.,][]{Boogaard2020,Aravena2014,Bothwell2013,Riechers2010}. The higher-$J$ CO transition observations (e.g., CO(5$-$4) and CO(6$-$5) probing warm and dense molecular gas) are necessary to establish whether the gas in the two SBs is highly excited as expected in starburst galaxies, or less excited as in MS galaxies.

\subsection{Dust mass}\label{dustmass}
To obtain the dust mass, $M_{\rm dust}$, we performed the infrared SED fitting with \texttt{CIGALE}\footnote{\texttt{CIGALE:} \url{https://cigale.lam.fr}} \citep[Code Investigating Galaxies Emission;][]{Burgarella2005,Noll2009,Boquien2019}. We fit data from 24 $\mu$m up to the millimeter wavelengths (see the data at 24 $\mu$m, 100 $\mu$m, 160 $\mu$m, 870 $\mu$m, and 1.8 mm in \citetalias{Wang2016}, and the data at 3.2 mm in Table \ref{tab1}). The dust infrared emission model \citep{Draine2014} combines two components with an extensive grid of parameters: the dust in the diffuse interstellar medium (ISM) heated by general diffuse starlight with a minimum radiation field $U_{\rm min}=0.1-50$ in the Habing unit \citep[1.2 $\times$ 10$^{-4}$ erg cm$^{-2}$  s$^{-1}$ sr$^{-1}$;][]{Habing1968}, and the dust tightly linked to star-forming regions illuminated by a variable radiation field ($U_{\rm min} < U < U_{\rm max}$) with a power-law distribution of a spectral index $\alpha=1-3$.  The fraction of dust mass linked to the star-forming regions is $\gamma=0.0001-1$. Among the two dust components, the mass fraction of dust in the form of the polycyclic aromatic hydrocarbons (PAH) is $q_{\rm PAH}=0.47-7.32 \%$. We note that a large grid of models was generated by \texttt{CIGALE} and then fitted to the observations to estimate physical properties; for more details, readers can refer to \cite{Boquien2019}. To investigate the active galactic nuclei (AGN) contributions to the infrared SEDs, we also fit the infrared SED with an additional AGN template \citep{Fritz2006} using CIGALE. We did not find any sign of a significant contribution of an AGN ($f_{\rm AGN}<0.2 \%$) in the infrared SEDs of our two SBs and two MSs, in agreement with previous literature constraints (\citetalias{Wang2016,Wang2018}). The derived $M_{\rm dust}$ is given in Table \ref{tab1}. It is used to calculate the gas mass (in $\S$\ref{Sec:Molecular Gas Mass}).

\begin{figure*}[h!]
        \begin{center}
        \includegraphics[width=0.99\textwidth]{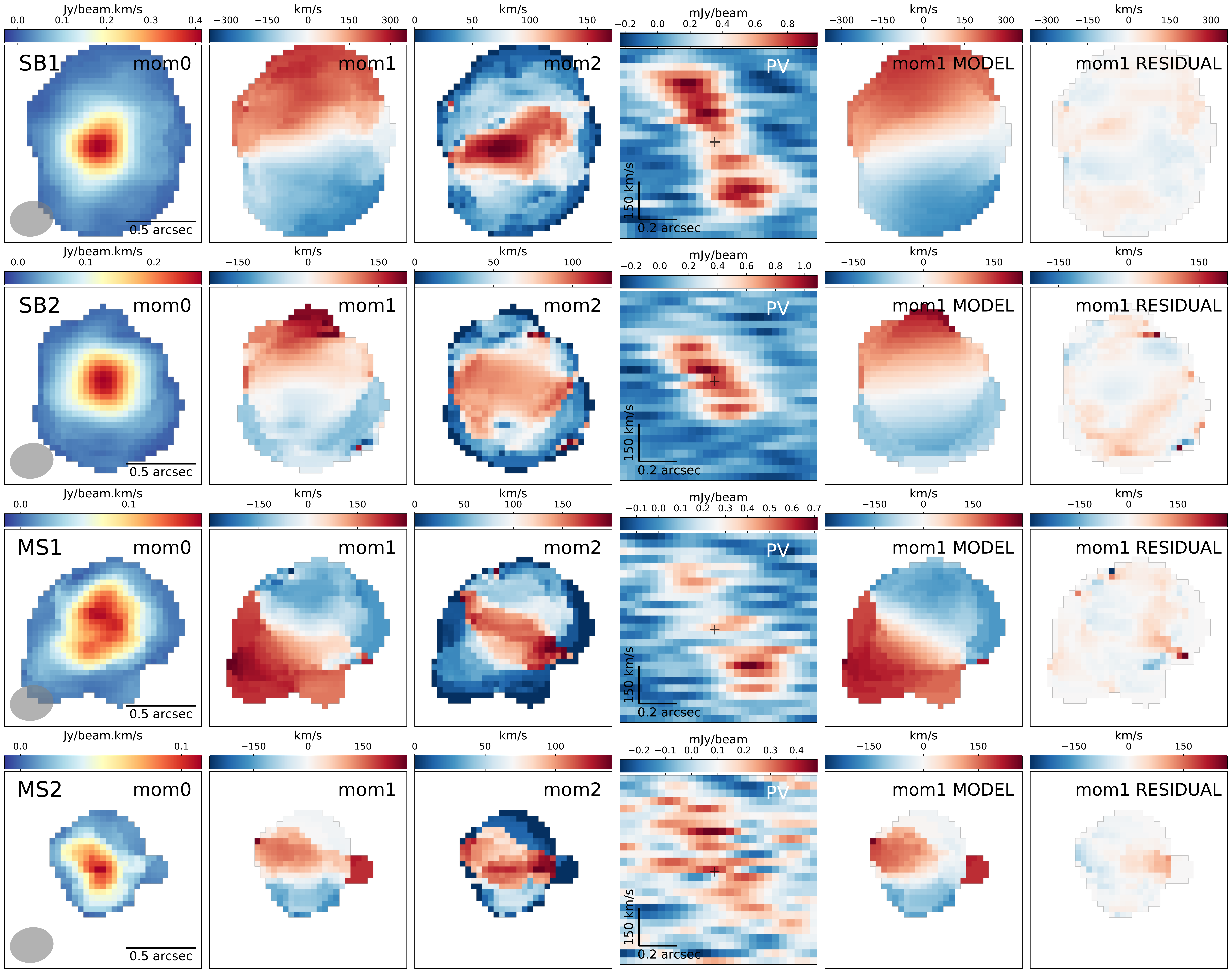} 
         \end{center}
          \vspace{-0.6truecm}  
        \caption{CO morphology and kinematics of the four cluster members at $z=2.51$. From left to right: ALMA maps ($1.4^{\prime\prime} \times 1.4^{\prime\prime}$) of velocity-integrated CO(1$-$0) flux ($Moment~0$), velocity field ($Moment~1$), velocity dispersion ($Moment~2$), position-velocity (PV) diagrams along the major axis, the best-fit $Moment~1$ model with GalPAK$^{3D}$, and the residual between the data and the model. We note that these maps are without correction for beam-smearing. Gray-filled ellipses indicate the angular resolution of $0.31^{\prime\prime} \times 0.25^{\prime\prime}$. The four member galaxies have regular rotating disks of molecular gas.   
        }
        \label{fig2}
        \vspace{-12pt}
\end{figure*}

\begin{figure*}[h!]
        \centering
        \includegraphics[width=1.\textwidth]{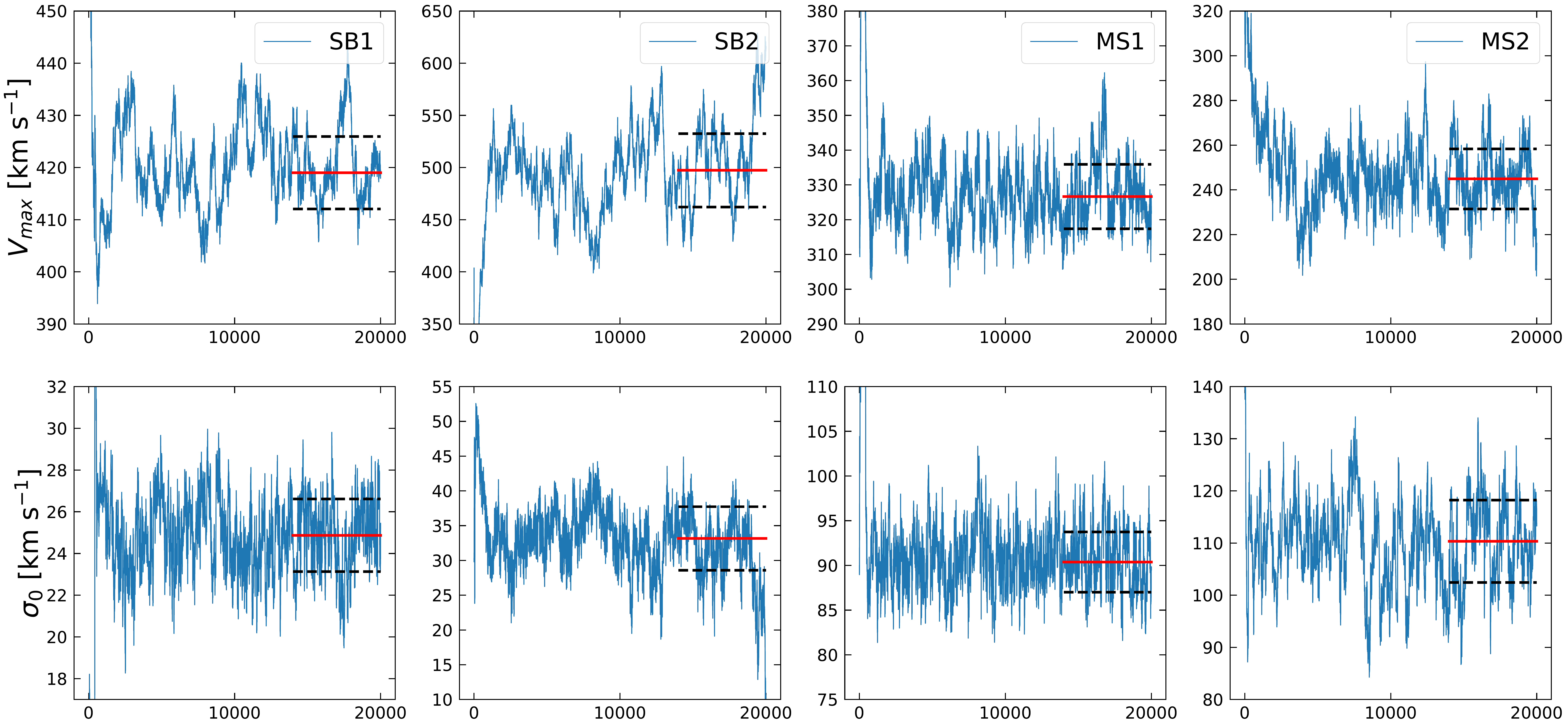}
        \caption{Full MCMC chain for 20,000 iterations for the two SBs and two MSs. Each galaxy shows the fitting results of rotational velocity and intrinsic velocity dispersion. Red solid lines and black dashed lines refer to the median and the 1$\sigma$ standard deviations of the last 30\% of the MCMC chain.}
        \label{ed1}
\end{figure*}

\begin{figure*}[h!]
        \centering
        \includegraphics[width=1\textwidth]{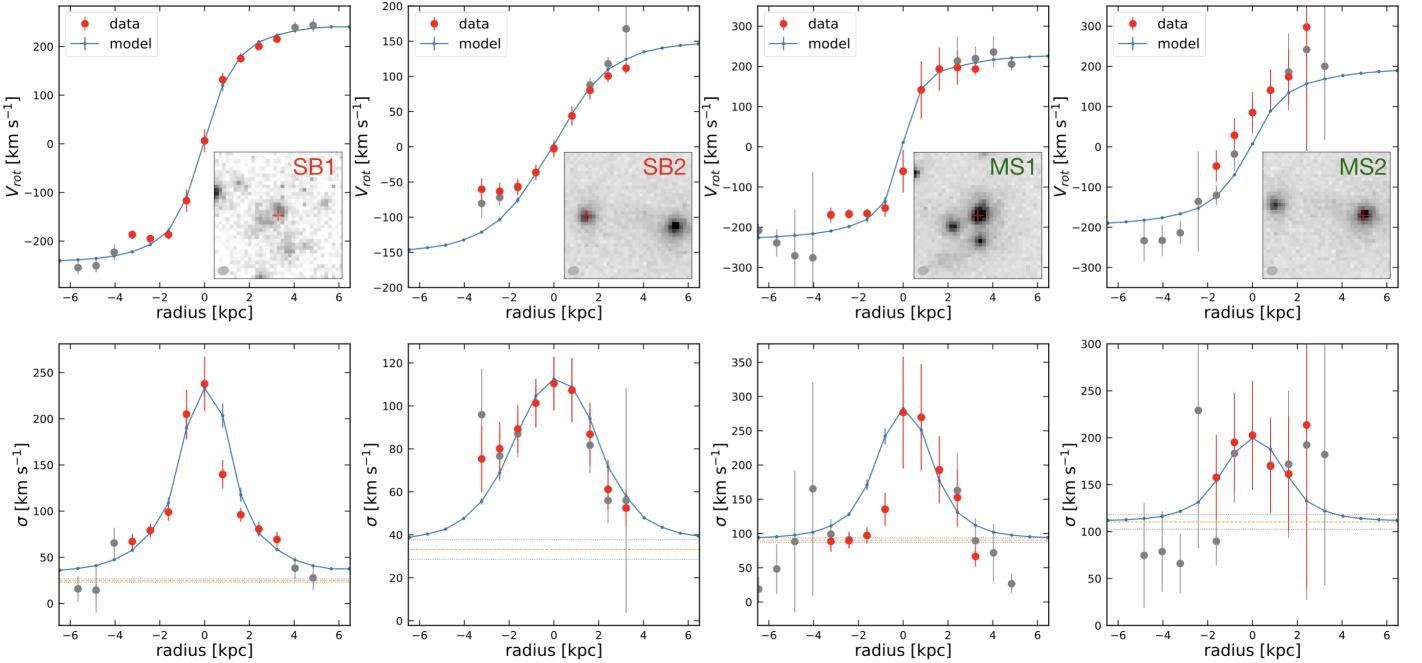}
        \caption{Observations (points) and best-fit PSF-convolved models (blue lines) for rotation velocity and total velocity dispersion profiles along the kinematic major axis. Each point was extracted from the spectrum in an aperture whose diameter is equal to the angular resolution 0.31$^{\prime\prime}$ and the red one restricted to the spaxels with a S/N $>$ 3 in the aperture.  Orange and gray lines mark the intrinsic velocity dispersion $\sigma_{\rm 0}$ and 1$\sigma$ confidence from the GalPAK$^{3D}$. The $3.5^{\prime\prime} \times 3.5^{\prime\prime}$ \textit{HST}/F160W images are shown in the bottom-right corner with the red crosses denote the corresponding source. Since SB2 and MS2 are close to each other with mixed fluxes, the points at the lowest radius of SB2 and the highest radius of MS2 are not fitted well with models. For MS1, the points at the lowest radius are not fitted well with models because of two neighboring sources at similar redshifts.  The $\sigma_{\rm 0}$ derived from GalPAK$^{3D}$ is lower than the lowest observed velocity dispersion profile because the latter one is still affected by the beam-smearing effect even at the large radii probed by our observations.}
        \label{ed2}
\end{figure*}

\begin{table*}\footnotesize  
\centering
        \caption{Summary of best-fit gas kinematic models for the two SBs and two MSs.} 
        
        \begin{tabular}{l c c c c} 
                \hline\hline
                \noalign{\smallskip}
                  & SB1 & SB2 & MS1 & MS2   \\
                \noalign{\medskip}
                \hline
                \noalign{\medskip}
                \multicolumn{1}{l} {Morphological parameters}& \\
                \noalign{\medskip}
                \hline
                \noalign{\smallskip}
                  $x_{\rm c}$ (pixel) & 16.53 $\pm$ 0.03 & 17.16 $\pm$ 0.04 & 15.70 $\pm$ 0.08 & 16.76 $\pm$ 0.14\\
                  $y_{\rm c}$ (pixel) & 16.67 $\pm$ 0.02 & 17.60 $\pm$ 0.03 & 16.73 $\pm$ 0.06 & 16.85 $\pm$ 0.10\\
                  $z_{\rm c}$ (pixel) & 12.75 $\pm$ 0.02 & 8.10 $\pm$ 0.02 & 10.22 $\pm$ 0.07 & 9.85 $\pm$ 0.13\\
                  Flux (Jy beam$^{-1}$) & 2.14 $\pm$ 0.01 & 1.17 $\pm$ 0.01 & 1.27 $\pm$ 0.02 & 0.67 $\pm$ 0.02\\
                  $r_{\rm 1/2}$ (kpc) & 1.79 $\pm$ 0.10 & 1.29 $\pm$ 0.09 & 2.43 $\pm$ 0.25 & 2.26 $\pm$ 1.04\\
                  $r_{\rm v}$ (kpc) & 0.17 $\pm$ 0.02 & 0.61 $\pm$ 0.03 & 0.02 $\pm$ 0.02 & 0.19 $\pm$ 0.08\\
                \hline
                \noalign{\medskip}
                \multicolumn{1}{c} {Orientation of gas disk:}& \\
                \noalign{\medskip}
                \hline
                  Incl. (deg) & 37.37 $\pm$ 0.62 & 19.30 $\pm$ 1.21 & 45.72 $\pm$ 1.38 & 55.58 $\pm$ 2.18\\
                  PA (deg) & 12.32 $\pm$ 0.32 & 9.94 $\pm$ 0.92 & 148.68 $\pm$ 0.96 & 47.75 $\pm$ 2.03\\
                \hline
                \noalign{\medskip}
                \multicolumn{1}{l} {Kinematic parameters$^{*}$}& \\
                \noalign{\medskip}
                \hline
                  $V_{\mathrm{max}}$ (km\,s$^{-1}$) & 419$^{+21}_{-58}$ & 497$^{+123}_{-252}$ & 327$^{+89}_{-47}$ & 245$^{+211}_{-97}$\\
                  $\sigma_{\mathrm{0}}$ (km\,s$^{-1}$)  & 25$^{+11}_{-15}$ & 33$^{+27}_{-19}$ & 90$^{+15}_{-18}$ & 110$^{+36}_{-40}$\\
                  $^{a} V_{\mathrm{max}}/\sigma_{\mathrm{0}}$ & 16.8$^{+7.4}_{-10.4}$ & 15.0$^{+13.0}_{-11.3}$ & 3.6$^{+1.1}_{-0.9}$ & 2.2$^{+2.0}_{-1.2}$ \\
                \hline
                \\
                \hline
                \noalign{\medskip}
                 reduced-$\chi^2$ & 1.26 & 1.50 & 1.12 & 1.01 \\
                \hline
                \multicolumn{5}{l}{$^{*}$ The uncertainties of the kinematic parameters were derived from simulations (see $\S$\ref{Molecular gas kinematics}).}\\
        \end{tabular}
        \tablefoot{Total ten free parameters and their 1$\sigma$ uncertainties were determined from the GalPAK$^{3D}$. The parameters are the $x_{\rm c}$, $y_{\rm c}$, and $z_{\rm c}$ positions; the total flux in the same units as the input data cube; the disk half-light radius $r_{\rm 1/2}$; the turnover radius $r_{\rm v}$; the inclination angle Incl. (the angle between the normal of the disk plane and the line-of-sight of the observer); position angle PA; the maximum circular velocity $V_{\mathrm{max}}$ ; and the intrinsic velocity dispersion $\sigma_{\mathrm{0}}$.  $^{a}$Rotation-to-random motions ratio, calculated from $V_{\mathrm{max}}$ and $\sigma_{\mathrm{0}}$. The bottom row is the reduced chi square $\chi^2$ in our MCMC fitting.}
        \label{summary}
\end{table*}

\subsection{Molecular gas mass}
\label{Sec:Molecular Gas Mass}
We adopted three commonly used methods to obtain the total molecular gas mass, $M_{\rm gas}$, and studied the gas properties of the two SBs and two MSs. The molecular gas masses were independently estimated based on the following: (i) the CO(1$-$0) line, (ii) the $M_{\rm dust}$ assuming a gas-to-dust mass ratio ($\delta_{\rm GDR}$), and (iii) the 3.2 mm dust continuum emission  (see Table \ref{tab3}).

To obtain the total molecular gas mass $M_{\rm gas,CO}$, we used the CO(1$-$0) luminosity (\citetalias{Wang2018}) instead of CO(3$-$2) to avoid uncertainties caused by the CO excitation ladder. The $M_{\rm gas,CO}$ was calculated by $M_{\rm gas,CO} = \alpha_{\rm CO}(Z) L_{\rm CO(1-0)}^{\prime}$, including a factor 1.36 correction for helium.
As was done in our previous work (\citetalias{Wang2018}), the metallicity-dependent CO-to-H$_2$ conversion factor, $\alpha_{\rm CO}(Z)$, was determined following \cite{Genzel2015} and \cite{Tacconi2018}. The metallicity was derived from the stellar mass, using the mass-metallicity relation \citep{Genzel2015}.

The gas mass, $M_{\rm gas,GDR}$, can also be determined through $M_{\rm dust}$ by employing the gas-to-dust ratio with a metallicity dependency \citep[e.g.,][]{Magdis2011,Magdis2012,Genzel2015,Bethermin2015}. They assumed that the gas-to-dust ratio is only related to the gas-phase metallicity ($\delta_{\rm GDR}-Z$) following the relation of \cite{Leroy2011}\footnote{Converted to the PP04 \citep{Pettini2004} metallicity scale by \cite{Magdis2012}.}:
\begin{flalign}
M_{\rm gas,GDR} =  \delta_{\rm GDR} M_{\rm dust},
\label{Mgas}
\end{flalign}
\begin{flalign}
\rm{log}\delta_{\rm GDR} = (10.54\pm1.0) - (0.99\pm0.12)\times(12 + \rm{log}(O/H)).
\label{GDR}
\end{flalign}
The metallicity was determined from the redshift-dependent mass-metallicity relation \citep[MZR;][]{Genzel2015}: 
\begin{flalign}
12 + \rm{log}(O/H) = a - 0.087(\rm{log}M_{\rm *} - b)^2, 
\label{Z}
\end{flalign}
where a = 8.74 and b = 10.4 + 4.46log(1$+z$)$-$1.78log(1$+z$)$^2$. The metallicity derived here is on the PP04 scale \citep{Pettini2004}, consistent with the calibration used in Eq.~\ref{GDR} by \cite{Magdis2012}. We adopted an uncertainty of 0.15 dex for the metallicities \citep{Magdis2012}. 

The long-wavelength Rayleigh-Jeans (RJ) tail ($\geq$$\sim$200 $\mu$m in the rest frame) of dust continuum emission is nearly always optically thin, providing a reliable probe of the total dust content. The RJ-tail method \citep{Scoville2016} is based on an empirical calibration between the molecular gas content and the dust continuum at the rest frame 850$\mu$m, as the fiducial wavelength, which was obtained after considering a sample of low-redshift star-forming galaxies, ultra-luminous infrared galaxies, and $z = 2-3$ submillimeter galaxies. We used our single ALMA 3.2 mm measurement at the RJ tail to derive the total molecular gas mass, $M_{\rm gas,3.2mm}$, following Eq.6 and Eq.16 (corrected using the published erratum) in \cite{Scoville2016}:
\begin{eqnarray}
\label{RJ}
M_{\rm gas,3.2mm}=1.78\, S_{\nu_{\rm obs}}{\rm [mJy]}\,(1+z)^{-(3+\beta)}\nonumber\\
\times \left( \frac{\nu_{\rm 850\mu m}}{\nu_{\rm obs}}\right)^{2+\beta}\,(d_{L}[{\rm Gpc}])^2\,\,\,\,\,\,\,\,\,\,\nonumber\\
\times \left\{ \frac{6.7\times10^{19}}{\alpha_{\rm 850}} \right\} \, \frac{\Gamma_{0}}{\Gamma_{\rm RJ}}\, 10^{10}M_{\odot}\,\,\nonumber\\
{\rm for} \, \, \lambda_{\rm rest} \gtrsim 250\, \mu {\rm m},\,\,\,\,\,\,\,\,\,\,\,\,\,\,\,\,\,\,\,\,\,\,\,\,\,\,
\end{eqnarray}
where $S_{\nu_{\rm obs}}$ is the observed flux density at the observed frequency $\nu_{\rm obs}$ corresponding to 3.2 mm, $\nu_{\rm 850\mu m}$ is the frequency corresponding to the rest-frame wavelength 850 $\mu m$, $d_L$ is the luminosity distance, and $\alpha_{850} = 6.7\times 10^{19}\,{\rm erg\, s^{-1}\, Hz^{-1}\,}M_{\odot}^{-1}$. 
The dust emissivity power-law spectral index $\beta$ is assumed to be 1.8. \ Furthermore, $\Gamma_{\rm RJ}$ is a correction for the deviation of the Planck function from RJ in the rest frame given by
\begin{equation}
\Gamma_{\rm RJ}(T_{\rm d}, \nu_{\rm obs}, z) = \frac{h\nu_{\mathrm obs}(1+z)/kT_{\rm d}}{e^{(h\nu_{\rm obs}(1+z)/kT_{\rm d})} - 1}, 
\end{equation}
\noindent
and $\Gamma_{0} = \Gamma_{\rm RJ}(T_{\rm d}, \nu_{850}, 0)$, where $h$ is the Planck constant and $k$ is the Boltzmann constant.
The mass-weighted dust temperature $T_d$ is assumed to be $25\ {\rm K}$, which is considered to be a representative value for both local star-forming galaxies and high-redshift galaxies \citep{Scoville2016}.

For our two SBs and two MSs, the $M_{\rm gas,CO}$, $M_{\rm gas,GDR}$, and $M_{\rm gas,3.2mm}$ derived from three different methods are consistent with each other (within one sigma for the two SBs and within two sigma for the two MSs).  All these values are presented in Table \ref{tab3}.

We note that in the empirical calibration of the RJ-tail method, $\alpha_{\rm CO}$ is set to 6.5 $M_\odot$ (K km s$^{-1}$ pc$^{2}$) $^{-1}$, which is different from what is used in our CO line method ($\alpha_{\rm CO}(Z)$ $\sim$ 4  $M_\odot$ (K km s$^{-1}$ pc$^{2}$) $^{-1}$ in Table \ref{tab3}), both of which include a correction factor of 1.36 for helium. 
To keep the independent measurements of $M_{\rm gas,3.2mm}$ and $M_{\rm gas,CO}$ and to maintain the consistency of the RJ-tail method commonly used in the literature, we did not perform any correction for $\alpha_{\rm CO}$ in the RJ-tail method. 

We also note that the derived $\alpha_{\rm CO}(Z)$ in \citetalias{Wang2018} is close to the Milky Way (MW) value (see Table \ref{tab3}).  In general, two conversion factors are commonly used in the literature to convert CO luminosities into gas masses, that is to say the MW\ ($\alpha_{\rm CO, MW}$ = 4.36 $M_{\odot}$(K km s$^{-1}$ pc$^{2}$)$^{-1}$) and local starbursts \citep[$\alpha_{\rm CO, SB}$ = 0.8 $M_{\odot}$(K km s$^{-1}$ pc$^{2}$)$^{-1}$;][]{Downes1998,Tacconi2008} values. However, using $\alpha_{\rm CO, SB}$ = 0.8 $M_{\odot}$(K km s$^{-1}$ pc$^{2}$)$^{-1}$ for the two SBs would result in very low gas-to-dust ratios ($\delta_{\rm GDR}$$\sim$20), making the two SBs extreme cases. The resulting $\delta_{\rm GDR}$ would be $4-5$ times lower than the typical $\delta_{\rm GDR}$ of solar-metallicity galaxies regardless of redshift \citep{Remy-Ruyer2014}, which is more than three times lower than that of GN20 \citep[$\delta_{\rm GDR}$$\sim$65 using $\alpha_{\rm CO}$ = 0.8;][]{Magdis2011,Magdis2012}, and about three times lower than the median $\delta_{\rm GDR}$ of local ultra-luminous infrared galaxies \citep{Solomon1997,Downes1998}. Therefore, we consider that using the derived $\alpha_{\rm CO}(Z)$ for the two SBs is more reasonable than using $\alpha_{\rm CO, SB}$ = 0.8 $M_{\odot}$(K km s$^{-1}$ pc$^{2}$)$^{-1}$.
In addition, some high-$z$ MS galaxies have been found to have gas properties of starbursts, the so-called SB in the MS, exhibiting compact dust structures \citep[e.g.,][]{Elbaz2018,Carlos2022}. Similarly, distant starburst galaxies can also exhibit MW-like ISM conditions when they present extended, instead of compact star formation \citep[e.g.,][]{Puglisi2021,Sharon2013} and/or large gas fractions. Here the two SBs exhibit both relatively extended CO sizes and large gas fractions (from the dust continuum and $\delta_{\rm GDR}$), favoring a MW-like $\alpha_{\rm CO}$. Hence, these pieces of evidence are all consistent with our derived $\alpha_{\rm CO}(Z)$. We however show that even if we had chosen a local SB conversion factor ($\alpha_{\rm CO, SB}$ = 0.8 $M_{\odot}$(K km s$^{-1}$ pc$^{2}$)$^{-1}$) for the two SBs, the main conclusions of the paper would remain unchanged. We emphasize that for the two SBs, the gas masses derived with the three independent methods are consistent within a 1$\sigma$ confidence level and the $M_{\rm gas,GDR}$ is independent of the assumption of the CO conversion factor.

It is worth mentioning that \cite{Gomez-Guijarro2019} also reported CO(1$-$0) luminosity values and \cite{Champagne2021} reported CO(1$-$0) luminosity and gas mass values for the two SBs and two MSs. We carefully compared our results with those derived using the CO(1$-$0) luminosity and/or gas mass from the above two papers, which mainly affect the gas fractions and Toomre Q values \citep[][described later in $\S$\ref{Disk stabilities}]{toomre}. We found that using the CO(1$-$0) luminosity (and gas mass) from different papers did not change our final results. They all agree with our results (gas fractions and Toomre Q values) within a 1$\sigma$ confidence level. 
We emphasize again that in our work, the gas masses are calculated using three different methods that are consistent with each other (within 1$\sigma$ confidence for the two SBs and within 2$\sigma$ for the two MSs). In particular, for the gas mass derived from $\delta_{\rm GDR}$, it is independent of the CO(1$-$0) luminosity (and $\alpha_{\rm CO}$), providing support to our results. 

 The derived molecular gas mass fractions, $f_{\rm gas} = M_{\mathrm{gas}}/(M_{\mathrm{*}} + M_{\mathrm{gas}})$, for the two SBs are $\sim$ 0.7 (0.70$^{+0.09}_{-0.06}$ and  0.66$^{+0.10}_{-0.07}$ based on the CO emission line, 0.68$^{+0.12}_{-0.10}$ and 0.65$^{+0.13}_{-0.11}$ based on the $\delta_{\rm GDR}$, and 0.73$^{+0.10}_{-0.08}$ and 0.61$^{+0.13}_{-0.11}$ based on the 3.2 mm dust continuum emission for the SB1 and SB2, respectively). They are two times lower for the two MSs,  $f_{\rm gas} \sim$ 0.3 (0.29$^{+0.09}_{-0.06}$ and 0.25$^{+0.08}_{-0.06}$ based on the CO emission line, 0.25$^{+0.15}_{-0.14}$ and 0.15$^{+0.08}_{-0.07}$ based on the $\delta_{\rm GDR}$, and 0.43$^{+0.13}_{-0.11}$ and 0.18$^{+0.08}_{-0.07}$ based on the 3.2 mm dust continuum emission for the MS1 and MS2, respectively). As a reference, the typical $f_{\rm gas}$ of $z=2.5$ SB and MS galaxies in the field with a stellar mass of $10^{11} M_\odot$ is 0.7 and 0.5, respectively \citep{Liu2019,Tacconi2018,Scoville2017}. It shows that the two cluster SBs are gas-rich, similarly to field SBs.

 \begin{figure*}[h!]
        \centering
        \includegraphics[scale=0.38]{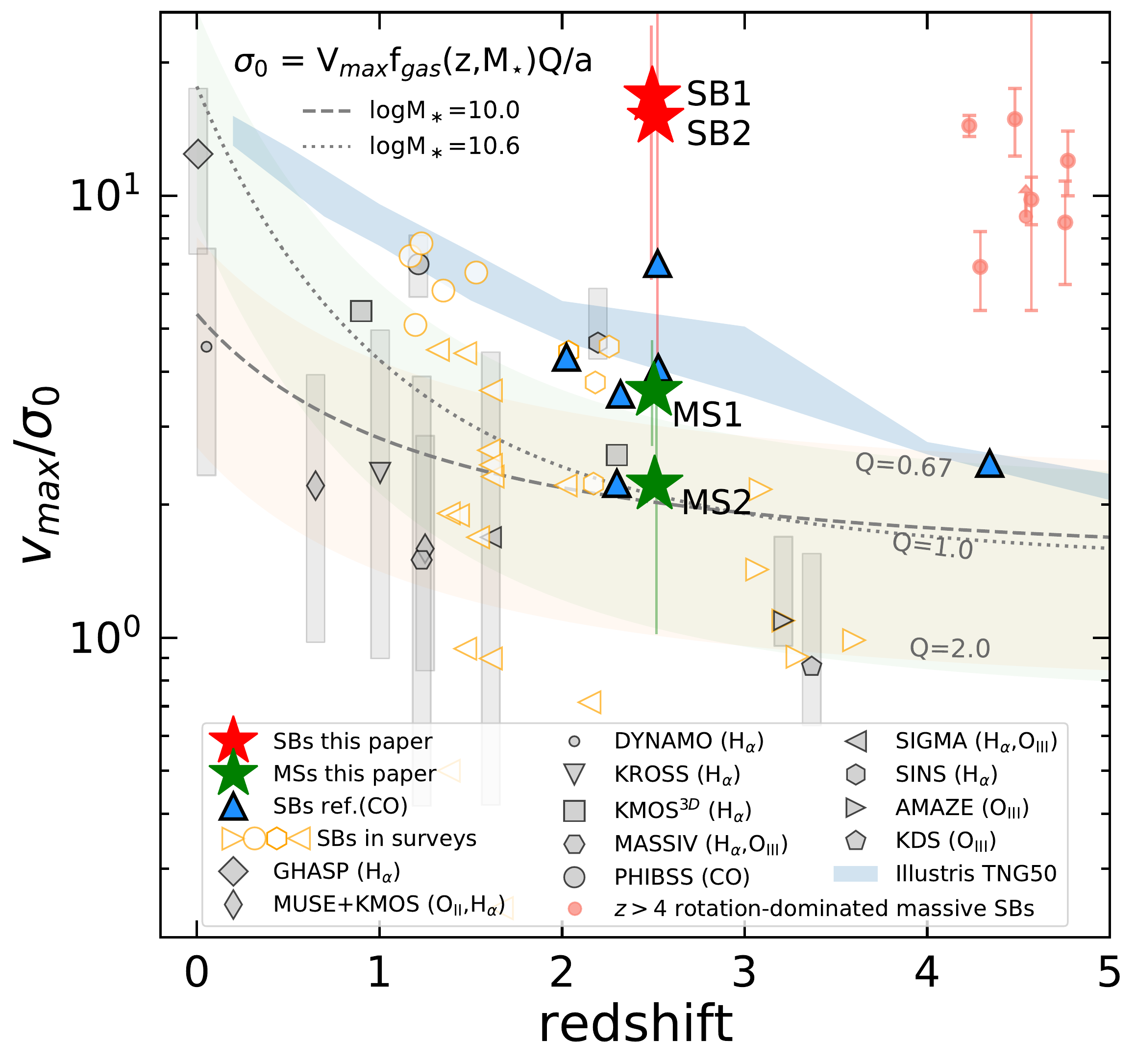}
        \includegraphics[scale=0.38]{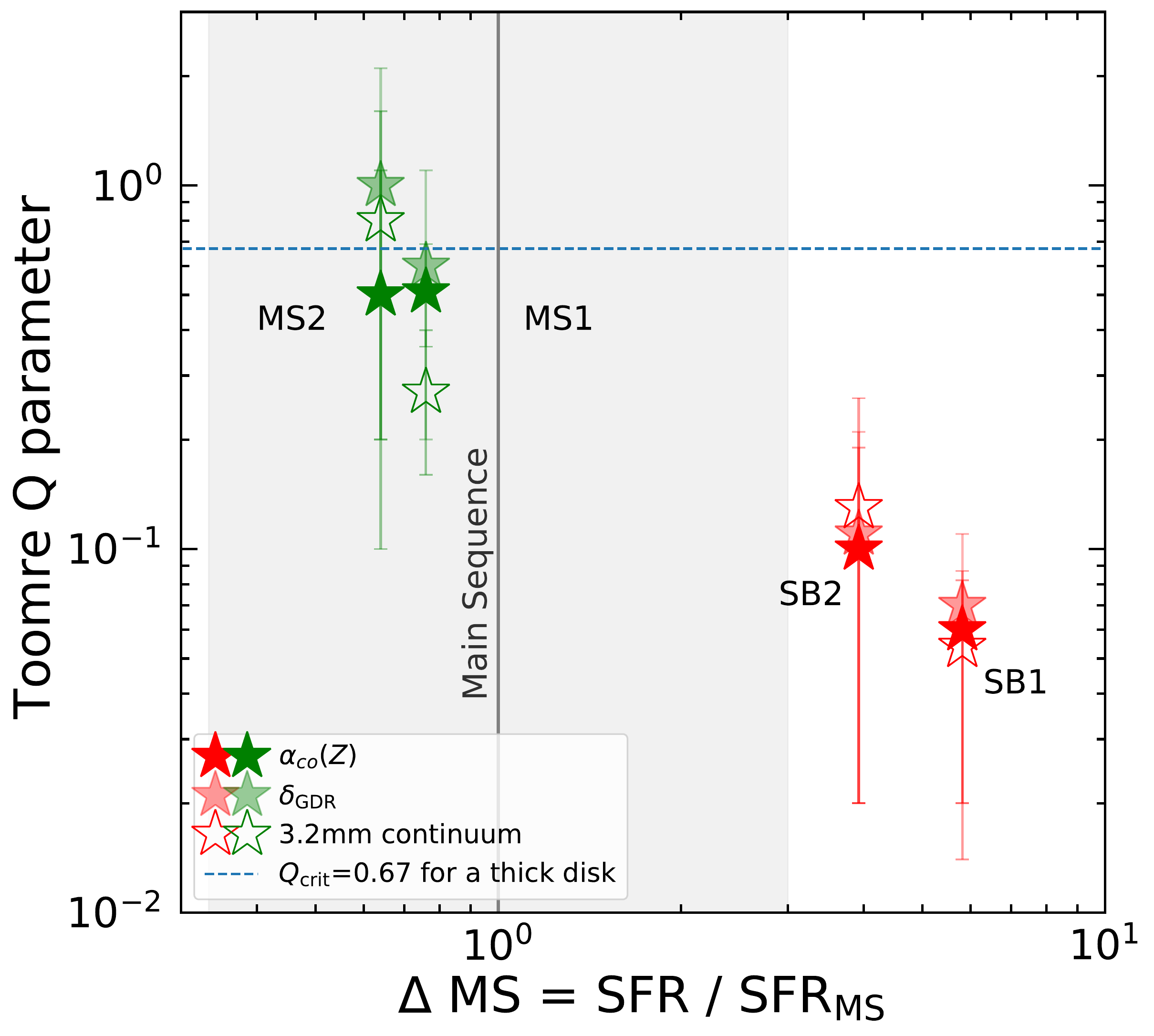}
        \caption{Two cluster SBs have dynamically cold and gravitationally unstable gas disks.
\textbf{Left}: Ratio of the gas rotational to random motion ($V_{\rm max}$/$\sigma_{\rm 0}$) as a function of redshift, with the comparison between the two cluster SBs and two cluster MSs and samples of observed and simulated field galaxies.  The two cluster SBs and two cluster MSs are in red and green stars, respectively, with uncertainties derived from our simulation (see $\S$\ref{Molecular gas kinematics}).
Filled gray symbols with vertical bars show the median values and 16-84th percentile range of field star-forming galaxies, including molecular and ionized gas detections. When possible, we identified field starbursts (orange symbols) within these literature samples at $z > 1$ as galaxies with a SFR at least 0.5 dex higher than the SFMS. 
 Blue triangles represent field starbursts with individual CO observations \citep{Rivera2018,Barro2017,Swinbank2011,Tadaki2017,Tadaki2019}. The faint red points show massive ($M_{\rm *} > 10^{10} M_\odot$) rotation-dominated SBs at $z > $ 4 which have been found recently \citep{Rizzo2020,Rizzo2021, Lelli2021, Fraternali2021}, but no information on their environment is available yet.
 The light-blue area shows $V_{\rm max}$/$\sigma_{\rm 0}$ values for star-forming galaxies from Illustris-TNG50 simulations in the mass range 10$^{9}$-10$^{11}$$M_{\odot}$  \citep{pillepich}. 
The two lines with the light green and pink shaded regions describe $V_{\mathrm{max}}/\sigma_{\mathrm{0}}$ as a function of $f_{\rm gas}$ and Toomre $Q$ \citep{Wisnioski2015}, where a = $\sqrt{2}$ for a disk with constant rotational velocity. 
\textbf{Right}: Toomre parameter $Q$ (see $\S$\ref{Disk stabilities}) as a function of the main-sequence offset. The solid gray line and shaded area highlight the SFMS \citep{Schreiber2015} position and the $\pm 3 \times \Delta$MS region, commonly used to separate MS from SB galaxies. We note that the two MSs have the same $\Delta$MS of 0.7 and are drawn staggered by 0.1 in this figure in order to distinguish between them.
The $Q$ values calculated using the 3.2 mm dust continuum, the $\delta_{\rm GDR}$, and the CO-based gas masses with $\alpha_{\rm CO}(Z)$ are shown by opened, filled light, and filled dark stars, respectively. 
The blue dashed line indicates the threshold $Q_{\rm crit}=0.67$ below which a thick disk (as assumed in our gas kinematic model) becomes gravitationally unstable.
}
        \label{fig3}
        \vspace{-12pt}
\end{figure*}

\subsection{Molecular gas kinematics}
\label{Molecular gas kinematics}
We investigated the kinematic properties of the molecular gas of the two cluster SBs and two cluster MSs by applying a three-dimensional (3D) kinematic modeling technique \citep[GalPAK$^{3D}$;][]{Bouche2015} to the CO(3$-$2) line cube. The 3D kinematic modeling technique is currently the most advanced kinematic modeling approach for the low resolution observations \cite[see e.g.,][]{Rivera2018, Fraternali2021, Tadaki2019, Fujimoto2021}. 
GalPAK$^{3D}$ is a Bayesian parametric tool, using a Markov Chain Monte Carlo (MCMC) approach to derive the intrinsic galaxy parameters and kinematic properties from a 3D data cube. 
The model is convolved with a 3D kernel to account for the instrument point spread function (PSF) and the line spread function (LSF) smearing effect. GalPAK$^{3D}$ can thus return intrinsic galaxy properties.
We adopted a thick exponential-disk light profile and an arctan rotational velocity curve to fit the 0.3$^{\prime\prime}$ resolution cube in order to determine the maximum circular velocity $V_{\rm max}$ and intrinsic velocity dispersion $\sigma_{\rm 0}$. 
The arctan rotational velocity curve is $V_{\rm rot} \propto V_{\rm max}$ arctan$(r/r_{\rm v})$, where $r_{\rm v}$ is the turnover radius.

Fig. \ref{fig2} shows the best-fit models and residuals, based on the last 30\% chain fraction with 20,000 iterations. For the two SBs and two MSs, our gas kinematic modeling provides a good description of the observed $Moment$ 0, 1, and 2 maps (Figs. \ref{res_SB1}, \ref{res_SB2}, \ref{res_MS1}, and \ref{res_MS2}). We also plotted the full MCMC chain from GalPAK$^{3D}$ for the two SBs and two MSs with 20,000 iterations to confirm the convergence of rotational velocity and intrinsic velocity dispersion (Fig. \ref{ed1}). The MCMC results of the fitting procedure are given in Figs. \ref{MCMC_SB1}, \ref{MCMC_SB2}, \ref{MCMC_MS1}, and \ref{MCMC_MS2}, showing the posterior probability distributions of the model parameters and their marginalized distributions.
The best-fitting values and their 1 $\sigma$ uncertainties are summarized in Table \ref{summary}.

To investigate the performance of the fit, we compared the rotation velocity and velocity dispersion profiles from the observed data with that from the best-fit PSF-convolved model data. All values were derived from one-dimensional (1D) spectra extracted using circular apertures with diameters equal to an angular resolution of 0.31$^{\prime\prime}$ along the kinematic major axis. As shown in Fig. \ref{ed2}, the profiles from dynamical models and observations are consistent. Some points deviate from the models because of the flux contamination by neighboring sources. The derived best-fit $\sigma_{\rm 0}$ from GalPAK$^{3D}$ is lower than the lowest observed velocity dispersion, which is reasonable because the latter is still affected by the beam-smearing effect even at the large radii probed by our observations.

Noticing that the best fit $\sigma_{\rm 0}$ of SB1 is lower than the CO(3$-$2) channel width of 30 km s$^{-1}$, we further tested the robustness of our fits. We applied the same analysis to the data cubes under the original channel width of $\sim$12 km s$^{-1}$. The results were consistent with each other within errors. Therefore, considering that a better sensitivity can help detect fainter galaxy edges and better sample  the flat part of the rotation curves, we have adopted the values from the data cubes with a channel width of 30 km s$^{-1}$ in this work.

Finally, we made a simple simulation to test the robustness of $\sigma_{\rm 0}$ and $V_{\rm max}$ given by the kinematic model of GalPAK$^{3D}$. Our purpose was to check first whether there is a global offset of the derived values or not, and second the reliability of the output uncertainties. To start, we randomly injected the best-fit PSF-convolved model data cube into the observed data cube (at the same positions as our galaxies, but at a different velocity and frequency to avoid the contamination of the CO emission from real sources). Then, we ran the GalPAK$^{3D}$ to derive $\sigma_{\rm 0}$ and $V_{\rm max}$. These two steps were repeated 100 times for each galaxy. In the end, we obtained distributions of $\sigma_{\rm 0}$ and $V_{\rm max}$. We calculated the median $\sigma_{\rm 0}$ and uncertainty (16-84th percentile range) values for the SB1, SB2, MS1, and MS2, respectively. 
In general, there is no global offset of the derived values from the GalPAK$^{3D}$, but the uncertainty is greatly underestimated by a factor of 5. Thus, we use the uncertainties of $\sigma_{\rm 0}$ and $V_{\rm max}$ given by this simulation in the main body of this paper, Fig. \ref{fig3}, Fig.~\ref{fig4}, and Table \ref{summary}.

\section{Results}
\label{Sec:results}
\subsection{The two SBs have dynamically cold gas-rich disks with low velocity dispersions}
\label{}
\begin{figure}[h!]
        \centering
        \includegraphics[scale=0.38]{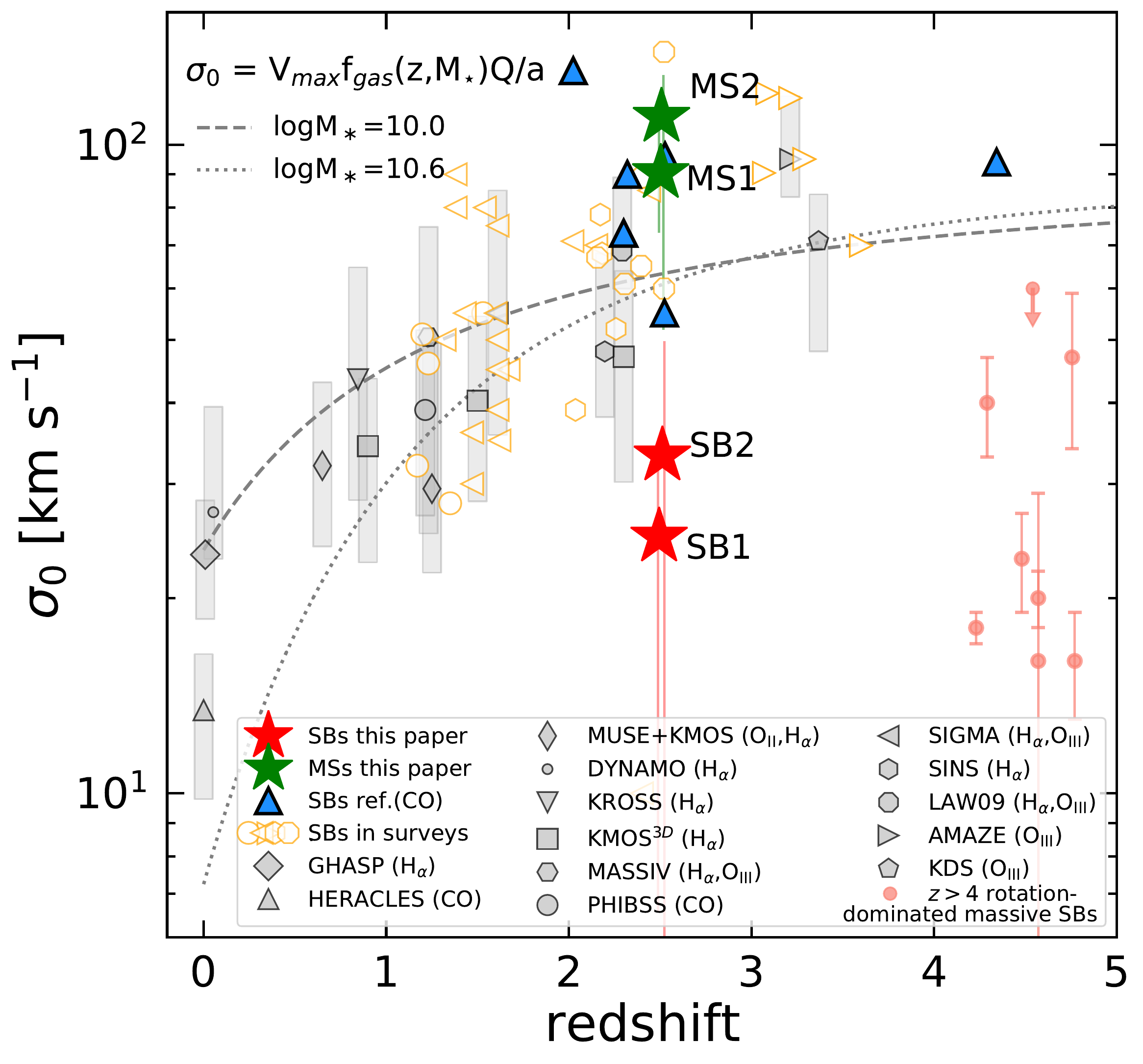}
        \caption{Comparison of gas turbulence between the two cluster SBs and two cluster MSs and field galaxies.
The intrinsic velocity dispersion increases with redshift. The symbol convention and the two lines are the same as in the left panel of Fig. \ref{fig3}. The two lines show the relation between the gas velocity dispersion, the gas fraction, and the disk instability for different stellar masses \citep{Wisnioski2015}. Here we assume $V_{\rm max}$ = 130 km s$^{-1}$, Q = 1, and a = $\sqrt{2}$ for a disk with constant rotational velocity. 
These tracks indicate that while $\sigma_0$ predictions from this relation depend on $M_{\rm *}$, this dependency mostly vanishes at $z>2$.  
The two cluster SBs have lower $\sigma_0$ than field SBs as well as most literature samples. Their uncertainties are derived from our simulation (see $\S$\ref{Molecular gas kinematics}).}
        \label{fig4}
        \vspace{-12pt}
\end{figure}

Our SBs and MSs are located in a dense environment. To better understand the environmental effect on gas turbulence, we compare the four cluster members with field galaxies from the literature. 
The data included in Fig. \ref{fig3} and Fig. \ref{fig4} contain molecular and ionized gas observations. 
The ionized gas observations of the star-forming galaxies, sorted by redshifts from the lowest to the highest, are from surveys GHASP \citep[log$M_{\rm *}$\lbrack$M_{\odot}$\rbrack=9.4-11.0; log$M_{\rm *}^{avg}$\lbrack$M_{\odot}$\rbrack=10.6;][]{Epinat2010}, DYNAMO \citep[log$M_{\rm *}$\lbrack$M_{\odot}$\rbrack=9.0-11.8; log$M_{\rm *}^{avg}$\lbrack$M_{\odot}$\rbrack=10.3;][]{Green2014},  MUSE and KMOS \citep[log$M_{\rm *}$\lbrack$M_{\odot}$\rbrack=8.0-11.1; log$M_{\rm *}^{avg}$\lbrack$M_{\odot}$\rbrack=9.4, 9.8 at $z$ $\sim$ 0.7, 1.3;][]{Swinbank2017}, KROSS \citep[log$M_{\rm *}$\lbrack$M_{\odot}$\rbrack=8.7-11.0; log$M_{\rm *}^{avg}$\lbrack$M_{\odot}$\rbrack=9.9;][]{Johnson2018}, KMOS$^{3D}$ \citep[log$M_{\rm *}$\lbrack$M_{\odot}$\rbrack=9.0-11.7; log$M_{\rm *}^{avg}$\lbrack$M_{\odot}$\rbrack=10.5, 10.6, 10.7 at $z$ $\sim$ 0.9, 1.5, 2.3;][]{Wisnioski2015,Ubler2019}, MASSIV \citep[ log$M_{\rm *}$\lbrack$M_{\odot}$\rbrack=9.4-11.0; log$M_{\rm *}^{avg}$\lbrack$M_{\odot}$\rbrack=10.2;][]{Epinat2012}, SIGMA \citep[log$M_{\rm *}$\lbrack$M_{\odot}$\rbrack=9.2-11.8; log$M_{\rm *}^{avg}$\lbrack$M_{\odot}$\rbrack=10.0;][]{Simons2016}, SINS \citep[log$M_{\rm *}$\lbrack$M_{\odot}$\rbrack=9.8-11.5; log$M_{\rm *}^{avg}$\lbrack$M_{\odot}$\rbrack=10.8;][]{Forster2009,Cresci2009}, LAW09 \citep[log$M_{\rm *}$\lbrack$M_{\odot}$\rbrack=9.0-10.9; log$M_{\rm *}^{avg}$\lbrack$M_{\odot}$\rbrack=10.3;][]{Law2009}, AMAZE \citep[log$M_{\rm *}$\lbrack$M_{\odot}$\rbrack=9.2-10.6; log$M_{\rm *}^{avg}$\lbrack$M_{\odot}$\rbrack=10.0;][]{Gnerucci2011}, and KDS \citep[log$M_{\rm *}$\lbrack$M_{\odot}$\rbrack=9.0-10.5; log$M_{\rm *}^{avg}$\lbrack$M_{\odot}$\rbrack=9.8;][]{Turner2017}. 
The molecular gas observations of the star-forming galaxies, sorted by redshifts, are from the HERACLES survey \citep[log$M_{\rm *}$\lbrack$M_{\odot}$\rbrack=7.1-10.9; log$M_{\rm *}^{avg}$\lbrack$M_{\odot}$\rbrack=10.5;][]{Leroy2008,Leroy2009} and the PHIBSS survey \citep[log$M_{\rm *}$\lbrack$M_{\odot}$\rbrack=10.6-11.2; log$M_{\rm *}^{avg}$\lbrack$M_{\odot}$\rbrack=11.0;][]{Tacconi2013}. We note that although the velocity dispersion measured from the molecular gas is $\sim$10$-$15 km s$^{-1}$ lower than from the ionized gas (with extra contributions from thermal broadening and expansion of the HII regions) in the local Universe, this difference becomes smaller with increasing redshift \citep{Ubler2019}. Some field SBs with individual CO observations \citep{Rivera2018,Barro2017,Swinbank2011,Tadaki2017,Tadaki2019} are indicated with blue triangles in Figs. \ref{fig3}  and \ref{fig4}. When possible, we also identified the field starbursts  (orange symbols in Figs. \ref{fig3} and \ref{fig4}) within the above star-forming samples, requiring their SFR to be at least 0.5 dex higher than the MS. 
In general, the gas in these field starbursts is slightly more turbulent than in main-sequence galaxies, showing a higher $\sigma_{\rm 0}$, especially at $2<z <4$. A similar trend of increasing $\sigma_{\rm 0}$ as galaxies move above the MS at a fixed stellar mass is shown for star-forming galaxies at $z=0-3$ \citep[e.g.,][]{Perna2022,Wisnioski2015}, suggesting that mergers or interactions would increase gas $\sigma_{\rm 0}$ as they enhance star formation.
 
 The gas disks of the two SBs are rotation-dominated with gas rotational to random motion ratios ($V_{\mathrm{max}}/\sigma_{\mathrm{0}}$) of 16.8$^{+7.4}_{-10.4}$ and 15.0$^{+13.0}_{-11.3}$, while the two MSs have a $V_{\mathrm{max}}/\sigma_{\mathrm{0}}$ of 3.6$^{+1.1}_{-0.9}$ and 2.2$^{+2.0}_{-1.2}$ (see Fig. \ref{fig3} and Table \ref{summary}). The $V_{\mathrm{max}}/\sigma_{\mathrm{0}}$ for the two SBs are more than three times higher than those for SBs and MSs in the field at the same epoch, either from observations or simulations. Moreover, they are as high as the median ratio for disk galaxies in the local Universe. As an example, the KMOS$^{3D}$ survey \citep{Wisnioski2015} found a median $V_{\mathrm{max}}/\sigma_{\mathrm{0}} \simeq 3$ \cite[$V_{\mathrm{max}}/\sigma_{\mathrm{0}} \sim 5$ in][]{Wisnioski2019} for field MS galaxies with stellar masses of $\sim$10$^{10.5}$$M_{\odot}$ at z $\sim$ 2.3. From Illustris-TNG50 simulations \citep{pillepich}, the typical star-forming galaxies in the stellar mass range 10$^{9}$$M_{\odot}$-10$^{11}$$M_{\odot}$ have $V_{\mathrm{max}}/\sigma_{\mathrm{0}}$ $\simeq 5$ $\pm$ 1 at $z=2.5$ (light-blue area in Fig. \ref{fig3}). In addition, using $V_{\mathrm{max}}/\sigma_{\mathrm{0}} \propto 1/f_{\rm gas}$($z$, $M_{\rm *}$) \citep{Wisnioski2015}, we get a value of $V_{\mathrm{max}}/\sigma_{\mathrm{0}} \simeq 2$ $\pm$ 1 at $z=2.5$, which decreases with increasing redshift (gray lines; light-green and pink-shaded regions in Fig. \ref{fig3}).

We note that there has recently been some debate about potential observational effects (beam-smearing effect) on the dynamical state of (field) high-$z$ galaxies \citep[e.g.,][]{Di Teodoro2016,Kohandel2020}. More specifically, when studying the gas kinematics with low spectral and spatial resolution data, the unresolved rotations within the PSF can artificially increase the value of $\sigma_{\mathrm{0}}$ and decrease the value of $V_{\mathrm{max}}$, leading to a severe systematic underestimation of the $V_{\mathrm{max}}/\sigma_{\mathrm{0}}$ ratio. The beam-smearing effect has been widely explored in local galaxies \citep[e.g., see Figure 6 in][] {Di Teodoro2015} and high-$z$ simulated galaxies \citep[e.g.,][]{Kohandel2020}. Subsequently, given the high spatial and spectral resolution of ALMA observations, which have been able to reach subkpc spatial resolution in high-$z$ galaxies, the beam-smearing effect is less significant. Compared to previous data, ALMA allows for a more robust way to measure intrinsic velocity dispersions. 
However, in our case, comparing two cluster SBs (with a spatial resolution of $\sim 0.3^{\prime\prime}$ and a channel width of 30 km s$^{-1}$) with other field SBs at similar redshifts and similar resolutions (with spatial resolutions of 0.1$^{\prime\prime}$-0.7$^{\prime\prime}$ and channel widths of 20-50 km s$^{-1}$, except for one with a channel width of $\sim$100 km s$^{-1}$), the $\sigma_{\rm 0}$ values of the two cluster SBs are about three times lower than that of those field SBs \citep[blue triangles in Fig. \ref{fig4};][]{Rivera2018,Barro2017,Swinbank2011,Tadaki2017}. 
In addition, even at the same spatial and spectroscopical resolution, the $\sigma_{\rm 0}$ values of the two cluster SBs are still more than three times lower than those of the two cluster MSs (green stars in Fig. \ref{fig4}).
Furthermore, the simulations have confirmed the robustness of $\sigma_{\mathrm{0}}$ and $V_{\mathrm{max}}$ for the two SBs and two MSs in our measurements (see $\S$\ref{Molecular gas kinematics}).
Therefore, we argue that the higher $V_{\mathrm{max}}/\sigma_{\mathrm{0}}$ and lower $\sigma_{\mathrm{0}}$ of the two cluster SBs with respect to field galaxies is a real physical difference, rather than being driven by observational effects.

In general, high-redshift star-forming galaxies are gas-rich and thus are believed to have a more turbulent ISM than local ones \citep{Forster-Schreiber2020}. While previous studies suggest that starburst galaxies are more likely associated with mergers or interactions, which are dynamically hot, the two cluster SBs surprisingly host dynamically cold disks, that is with a large $V_{\mathrm{max}}/\sigma_{\mathrm{0}}$.
Recently, observations also found this type of phenomenon at $z > $ 4 with the presence of massive ($M_{\rm *} > 10^{10} M_\odot$) rotation-dominated SBs \citep[faint red points in Fig. \ref{fig3};][]{Rizzo2020,Rizzo2021, Lelli2021, Fraternali2021}, but no information on their environment is available yet.

The rotationally supported and dynamically cold disks of the two SBs imply low gas turbulent motions that are lower than for field SBs and MSs at the same redshift (Fig. \ref{fig4}). Hence, the two cluster SBs appear to be weakly affected by extreme  internal and$/$or external physical processes, such as stellar or AGN feedback and galaxy mergers. 

\subsection{The two SBs have gravitationally unstable gas disks}
\label{Disk stabilities}
To understand the triggering mechanism taking place in these two SBs, we further study the dynamical state of their disks.  A rotating, symmetric gas disk is unstable with respect to the gravitational fragmentation if the Toomre $Q$ parameter, $Q=\kappa$$\sigma_{\rm 0,gas}$/$(\pi G{\Sigma_{\rm gas}})$ \citep{toomre}, is below a threshold value of $Q_{\rm crit}$. The $Q_{\rm crit}=1$ is for a thin gas disk, and $Q_{\rm crit}=0.67$ for a thick gas disk. We calculated the gas surface density, ${\Sigma_{\rm gas}} = 0.5M_{\rm gas}/(\pi r_{\rm 1/2}^2)$, where $M_{\rm gas}$ was derived from the CO(1$-$0) luminosity, the $M_{\rm dust}$ assuming a gas-to-dust mass ratio, and dust continuum emission at 3.2 mm.
Here we adopt the maximum circular velocity $V_{\rm max}$, intrinsic velocity dispersion $\sigma_{\rm 0}$, the disk half-light radius $r_{\rm 1/2}$, and flat rotation curve with the epicyclic frequency $\kappa$=$\sqrt{2}$$V_{\rm max}/r_{\rm 1/2}$ \citep{Binney2008}. All of these values were derived from the gas kinematic modeling. 

The right panel of Fig. \ref{fig3} shows the $Q$ values of the two SBs and two MSs based on CO, $\delta_{\rm GDR}$, and dust continuum emission as a function of the main-sequence offset.
For the two SBs and two MSs, the derived $Q$ values decrease as the SFR offset to the SFMS increases. The two SBs present $Q$ values much smaller than $Q_{\rm crit}$. We note that using the local starbursts' conversion factor ($\alpha_{\rm CO, SB}$ = 0.8 $M_{\odot}$(K km s$^{-1}$ pc$^{2}$)$^{-1}$) for the two SBs, the derived Toomre parameter $Q$ values increase by a factor of five but still remain below $Q_{\rm crit}$. The low Toomre $Q$ means that the gas disk can easily collapse due to gravitational instabilities, leading to a starburst (see $\S$\ref{Triggering mechanism of high-$z$ cluster starburst} for a more in-depth discussion). This is consistent with the fact that the SFE (1/$t_{\rm dep}$) values of the two SBs are $\gg$ 0.5 dex than the scaling relation of MS galaxies \citep{Liu2019}, suggesting that they have efficient star formation. 

\section{Discussion}
\label{Discussion}
\subsection{Triggering mechanism of high-$z$ cluster starburst}
\label{Triggering mechanism of high-$z$ cluster starburst}

The dynamically cold, rotating gas disks of the two SBs are mainly due to a low level of gas turbulent motions. Their intrinsic velocity dispersions are extremely low, with $\sigma_{\rm 0} \sim 20-30$ km\,s$^{-1}$, while the two MSs have $\sigma_{\rm 0} \sim 90-110$ km\,s$^{-1}$ (see Table \ref{summary}). Their $\sigma_{\rm 0}$ values are about three times lower than that of field SBs at similar redshifts
 based on molecular gas observations from 2D and 3D kinematic modeling \citep[blue triangles in Fig. \ref{fig4};][]{Rivera2018,Barro2017,Swinbank2011,Tadaki2017}, and even field MS galaxies.
We note that the $\sigma_{\rm 0}$ values of the two cluster SBs are consistent with those of a group of massive ($M_{\rm *} > 10^{10} M_\odot$) rotation-dominated SBs recently found at $z > 4$ \citep[faint red points in Fig. \ref{fig4};][]{Rizzo2020,Rizzo2021, Lelli2021, Fraternali2021}. 
In general, the gas in field SBs at $z = 2-3$ (blue and orange markers in Fig. \ref{fig4}) seems slightly more turbulent than that in field MSs, with a higher median $\sigma_{\rm 0}$ by a factor of $\sim$1.5. Hence, the two cluster SBs exhibit strikingly different kinematic properties compared to field SBs at $z$ $\sim$ 2.5, whether they are observed by molecular gas or ionized gas. 
These results challenge our current understanding of the physical origin of high-$z$ starbursts in clusters and hint that mergers and interactions are not the only way to trigger starbursts. 

In addition, in the study of gas disk stabilities (see $\S$\ref{Disk stabilities} and Fig. \ref{fig3}), we find the two SBs have unstable gas disks, as shown by their relatively low Toomre $Q$ values. This means that their gas disks can easily collapse due to gravitational instabilities. According to current theories of galaxy evolution, the self-regulated star formation of galaxies require marginally gravitationally unstable disks \citep{Thompson2005,Cacciato2012}. In short, the accretion of gas from the intergalactic medium to a gas disk increases the self-gravity of the gas disk. Once the self-gravity of the gas overcomes stellar radiation pressure, the gas disk becomes unstable and can easily collapse, triggering star-formation. As the stellar feedback injects energy into the ISM, thus increasing gas turbulence and radiation pressure, the gas disk then becomes stable. Subsequently, star formation becomes inefficient and the gas self-gravity dominates again. Therefore, the Toomre $Q$ value of such a marginally unstable disk is always around $Q_{crit}$. However, in our case, the two SBs have much lower Toomre $Q$ values than $Q_{\rm crit}$ and they have high gas fractions ($\sim$0.7; see Table \ref{tab3}). The gas turbulence in these two SBs is very inefficient in balancing the gas self-gravity, leading to starbursts. 
In other words, in our work, the combination of the high gas fractions and the low velocity dispersions of the two SBs would naturally yield highly unstable gas disks, which in turn induce efficient starburst activities (Fig. \ref{fig3}). 
In fact, the two SBs have higher gas fractions than other MS galaxies, but they are still comparable to other SBs in the field (see $\S$\ref{Sec:Molecular Gas Mass}). It is their low velocity dispersions that make the two cluster SBs unusual. 

Thus, the most possible mechanism for inducing efficient star formation activity in the two cluster SBs is the self-gravitational instability of gas disks caused by the low $\sigma_{0}$ (and the high gas fractions). 
In the following section, we discuss possible scenarios most likely to lead to the low $\sigma_{0}$ and high gas fractions, which are believed to be the coplanar, corotating cold gas accretion through the cold cosmic-web streams.

\subsection{Drivers of low turbulent gas in the two cluster SBs}
\label{Sec:Low Turbulent Gas in The Two SBs}

We further explore the physical origin of the extremely low velocity dispersions and high gas fractions of the cluster SBs.  The two primary energy sources of gas turbulence are stellar feedback and gravitational instability \citep{Krumholz2016,Krumholz2018}, while a high gas fraction could either be due to a merger or gas infall (cold gas accretion). The two SBs favor a scenario in which cold gas is accreted via corotating and coplanar streams \citep[e.g.,][]{Danovich2012,Danovich2015}. Indeed, in such a scenario, most of the gravitational energy is converted to rotation rather than dispersion, thus building up angular momentum \citep{Kretschmer2020,Kretschmer2021}. 
 As a result, the two cluster SBs would have high gas fractions, low turbulent gas disks with high rotational velocities ($V_{\mathrm{max}} \sim 400 - 500$ km\,s$^{-1}$; see Table \ref{summary}), and relatively extended gas disk sizes. 
Simulations do predict a high occurrence rate of such configurations of corotating and coplanar cold gas accretion in massive galaxies and halos \citep{Kretschmer2021,Dekel2020a} for which galaxy merger events are rare enough to allow gas disks to survive over a long timescale. 
More specifically, the cold gas disk with $V_{\mathrm{max}}/\sigma_{\mathrm{0}} \simeq 5$ in a $z\sim3.5$ galaxy, with a stellar mass of $M_{\rm *} \sim 10^{10} M_\odot$, typically survives for approximately five orbital periods (a duration time of $\sim410$ Myr) \citep{Kretschmer2021}, before being disrupted by a merger. Assuming the same duration time of 410 Myr between two merger events, we infer that our two cluster SBs, which have more massive cold gas disks with larger $V_{\mathrm{max}}/\sigma_{\mathrm{0}} > 10$, would survive for $\gtrsim15$ orbital periods to mergers. 

Gas-rich major mergers can also lead to high gas fractions in the merger remnants. Numerical simulations and observations further indicate that rotating disks can reform rapidly after the final coalescence stage of the gas-rich mergers \citep[e.g.,][]{Robertson2006,Hopkins2009,Ueda2014}. 
However, simulations of gas-rich major mergers reveal enhanced gas turbulence and reduced sizes of disk components that survive or that one rebuilt after a merger \citep[e.g.,][]{Bournaud2011}, which conflicts with the observed low $\sigma_{\mathrm{0}}$ and relatively extended disk sizes of the two SBs. We highlight that the $\sigma_{\mathrm{0}}$ values of the two SBs are significantly lower than their field counterparts (not only field SBs, but also most field MSs). 
Meanwhile, we have not seen clear evidence of galaxy mergers in the two cluster SBs; this not only includes the CO(3$-$2) map, but also the high-resolution HST/F160W image tracing the stellar structures. While the possibility cannot
be fully ruled out that the two SBs are in the final coalescence stage of gas-rich mergers, the chance that both SBs are caught in this short stage is small.  Thus, the high gas fraction,  low $\sigma_{0}$, and lack of evidence for galaxy major mergers in the two SBs appear inconsistent with merger remnants. 

In summary, the two cluster SBs have suppressed velocity dispersion and high gas fraction. These two unique properties of the two SBs support the scenario of corotating and coplanar cold gas accretion as the cause of their highly efficient star formation.

\section{Conclusions}
\label{Sec:summary}

In this paper, we have studied the molecular gas kinematics of two SBs and two MSs in the most distant known X-ray cluster CLJ1001 at $z=2.51$ (Fig.~\ref{fig1}), based on ALMA high-resolution CO(3$-$2) observations ($\sim$0.3$^{\prime\prime}$ corresponding to $\sim$2.5 kpc). These four galaxies show regular rotating gas disks, without clear evidence of a past gas-rich major merger (Fig.~\ref{fig2}). While exploring the disk stabilities, we find strong evidence that the two cluster SBs have gravitationally unstable gas disks (right panel of Fig.~\ref{fig3}). Therefore, we suggest that self-gravitational instability of the gas disks is the most likely mechanism that induces intense  star formation in the two cluster SBs.

The two cluster SBs show dynamically cold gas-rich disks with significantly higher $V_{\mathrm{max}}/\sigma_{\mathrm{0}}$ than their field counterparts at similar redshifts, implying that their gas is low-turbulent (left panel of Fig.~\ref{fig3} and Fig.~\ref{fig4}). The gas disks of the two cluster SBs have extremely low velocity dispersions ($\sigma_{\mathrm{0}} \sim 20-30$ km s$^{-1}$), which are three times lower than their field counterparts at similar redshifts.
The suppressed velocity dispersions (see Table \ref{summary}) and high gas fractions ($\sim$0.7; see Table \ref{tab3}) of the two cluster SBs yield gravitationally unstable gas disks, which enable highly efficient star formation. The unique properties of the two cluster SBs (high gas fraction and suppressed velocity dispersion) support the scenario in which corotating and coplanar cold gas accretion might serve as an essential avenue to trigger starbursts in forming galaxy clusters at high redshift. This may represent an important process other than mergers and interactions in triggering starbursts at high redshifts, at least in massive halos. Characterizing the environment of the recently found massive ($M_{\rm *} > 10^{10} M_\odot$) rotation-dominated disks at $z > $ 4 with suppressed velocity dispersion presents a unique opportunity to support this scenario further.

\begin{acknowledgements}
We are very grateful to the anonymous referee for instructive comments, which helped to improve the overall quality and strengthened the analyses of this work. We thank Ken-ichi Tadaki and Nicolas Bouch{\'e} for the helpful discussion of the molecular gas kinematics with GalPAK$^{3D}$; Yong Shi, Alain Omont, Susan Kassin, Annagrazia Puglisi, Daizhong Liu, and Boris Sindhu Kalita for valuable discussions and suggestions that improved this paper. This paper makes use of the following ALMA data: ADS/JAO.ALMA \#2016.1.01155.S. ALMA is a partnership of ESO (representing its member states), NSF (USA), and NINS (Japan), together with NRC (Canada), NSC, ASIAA (Taiwan), and KASI (Republic of Korea), in cooperation with the Republic of Chile. The Joint ALMA Observatory is operated by ESO, AUI/NRAO, and NAOJ. 
This work is supported by the National Key Research and Development Program of China (No. 2017YFA0402703), and by the National Natural Science Foundation of China (Key Project No. 11733002, Project No. 12173017, and Project No. 12141301).
This work is also supported by the Programme National Cosmology et Galaxies (PNCG) of CNRS/INSU with INP and IN2P3, co-funded by CEA and CNES.
M.-Y. X. acknowledges the support by China Scholarship Council (CSC). T.W. and K. K. acknowledges the support by NAOJ ALMA Scientific Research Grant Number 2017-06B. 
D. I. acknowledges the support by JSPS KAKENHI Grant Number JP21H01133. 
     
\end{acknowledgements}

\onecolumn
\begin{appendix}
\section{Gas Kinematics: Comparison between the data and the best-fit model}

\begin{figure*}[h!]
\centering
        \includegraphics[scale=0.37]{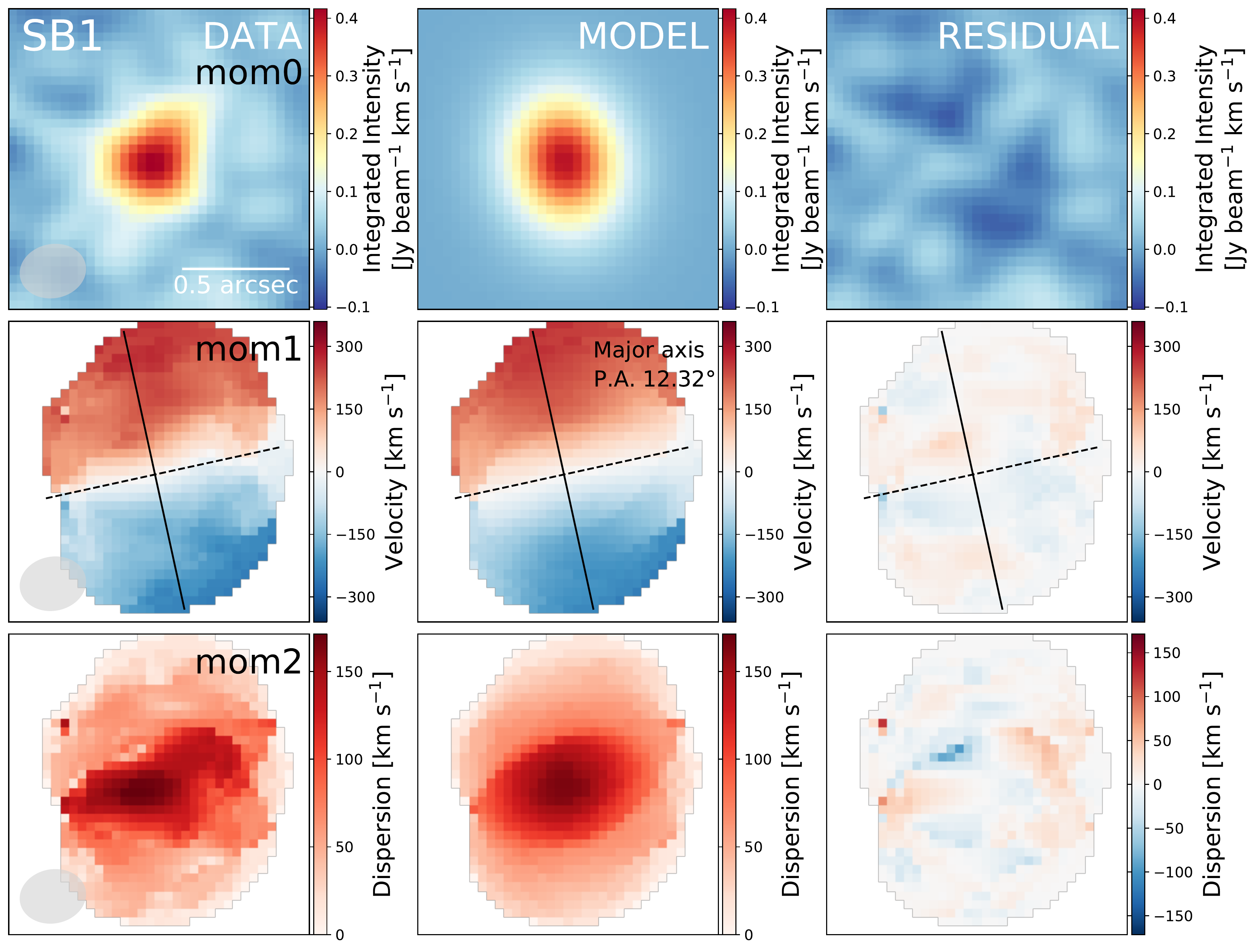}
        \caption{Comparison between the data and the best-fit model for the SB1. From top to bottom, this figure shows velocity-integrated CO(3$-$2) flux density ($Moment~0$), velocity field ($Moment~1$), and velocity dispersion ($Moment~2$) maps for the data (left panel); the best-fit model (middle panel); and the residual after subtracting the model from the data (right panel).  Gray-filled ellipses in the bottom-left corner indicate the angular resolution of $0.31^{\prime\prime} \times 0.25^{\prime\prime}$. Each panel is $1.4^{\prime\prime} \times 1.4^{\prime\prime}$, with the north being up and the east to the left.
}
        \label{res_SB1}
\end{figure*}

\begin{figure*}[h!]
        \centering
        \includegraphics[scale=0.37]{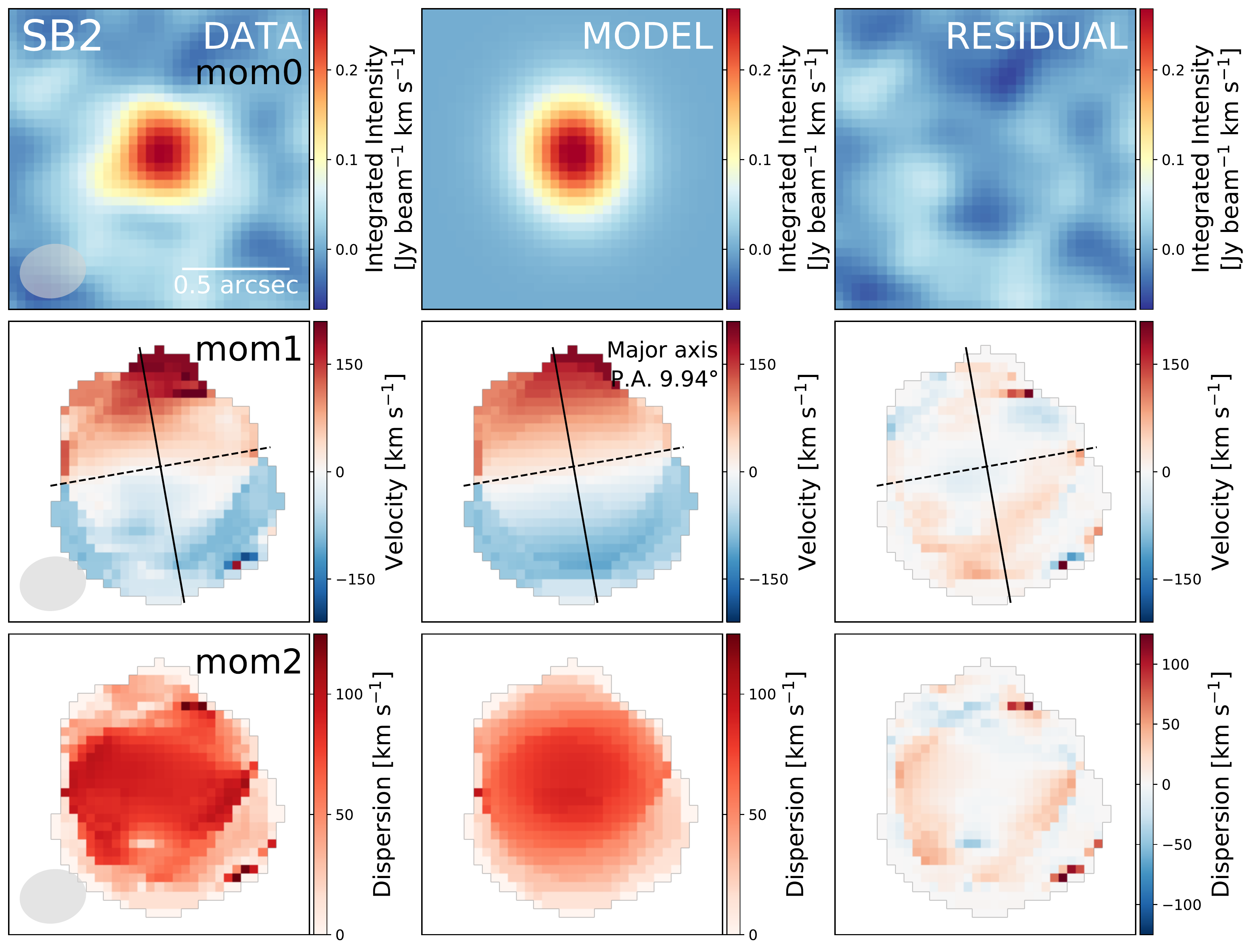}
        \caption{Same as Fig. \ref{res_SB1}, but for the SB2.} 
        \label{res_SB2}
\end{figure*}

\begin{figure*}[h!]
        \centering
        \includegraphics[scale=0.37]{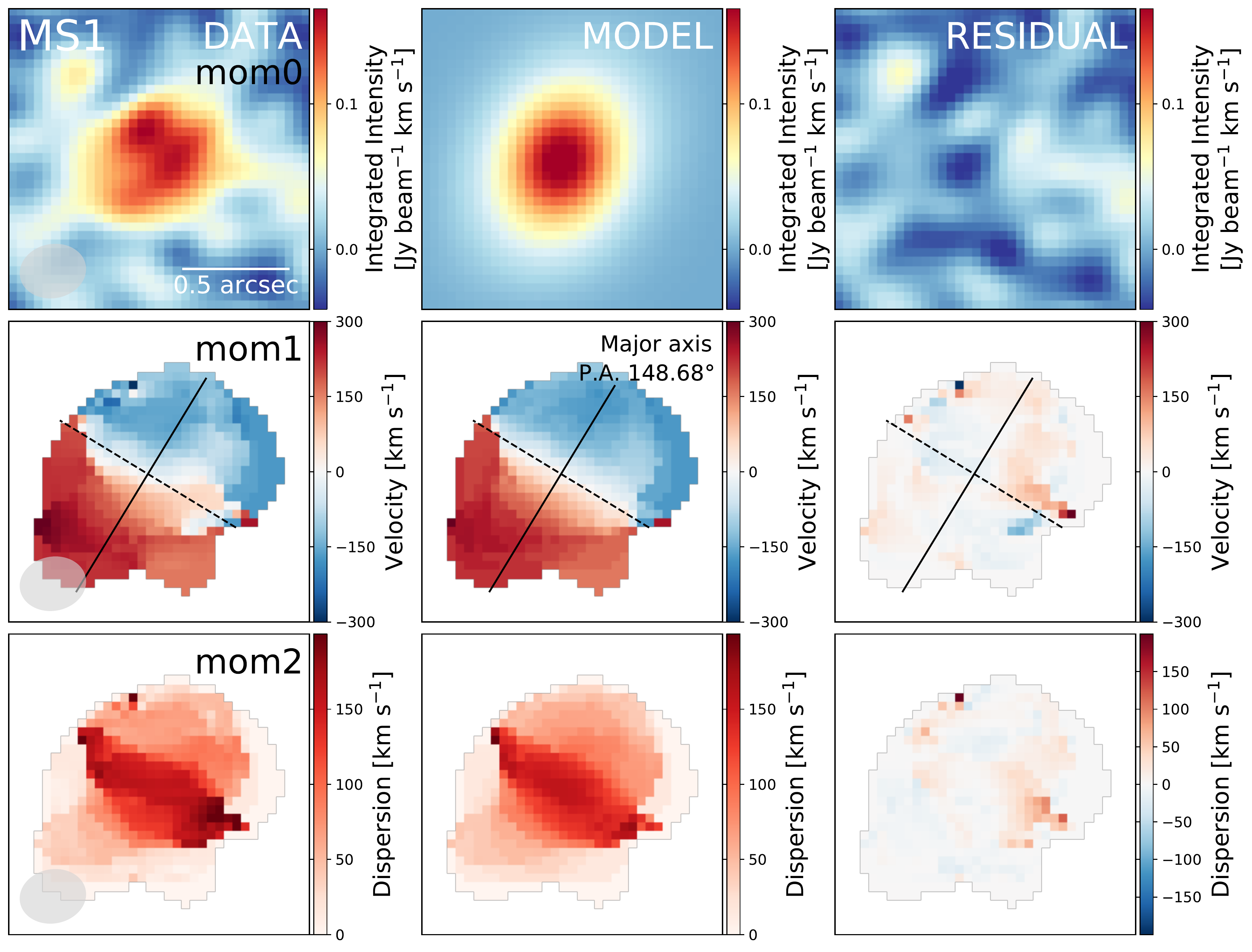}
        \caption{Same as Fig. \ref{res_SB1}, but for the MS1.} 
        \label{res_MS1}
\end{figure*}

\begin{figure*}[h!]
        \centering
        \includegraphics[scale=0.37]{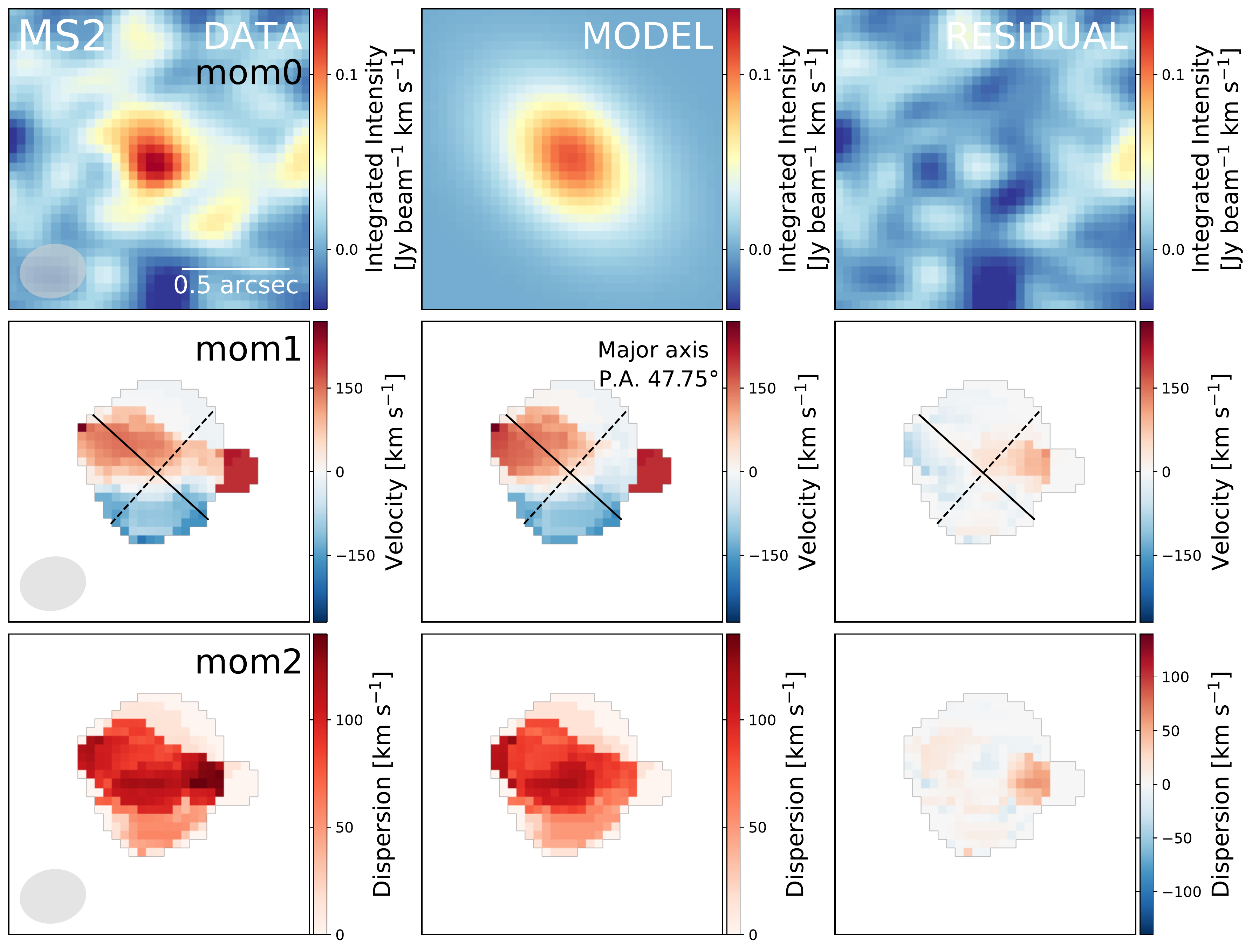}
        \caption{Same as Fig. \ref{res_SB1}, but for the MS2.} 
        \label{res_MS2}
\end{figure*}
\onecolumn
\section{Gas Kinematics: MCMC results}

\begin{figure*}[h!]
        \centering
        \includegraphics[scale=0.54]{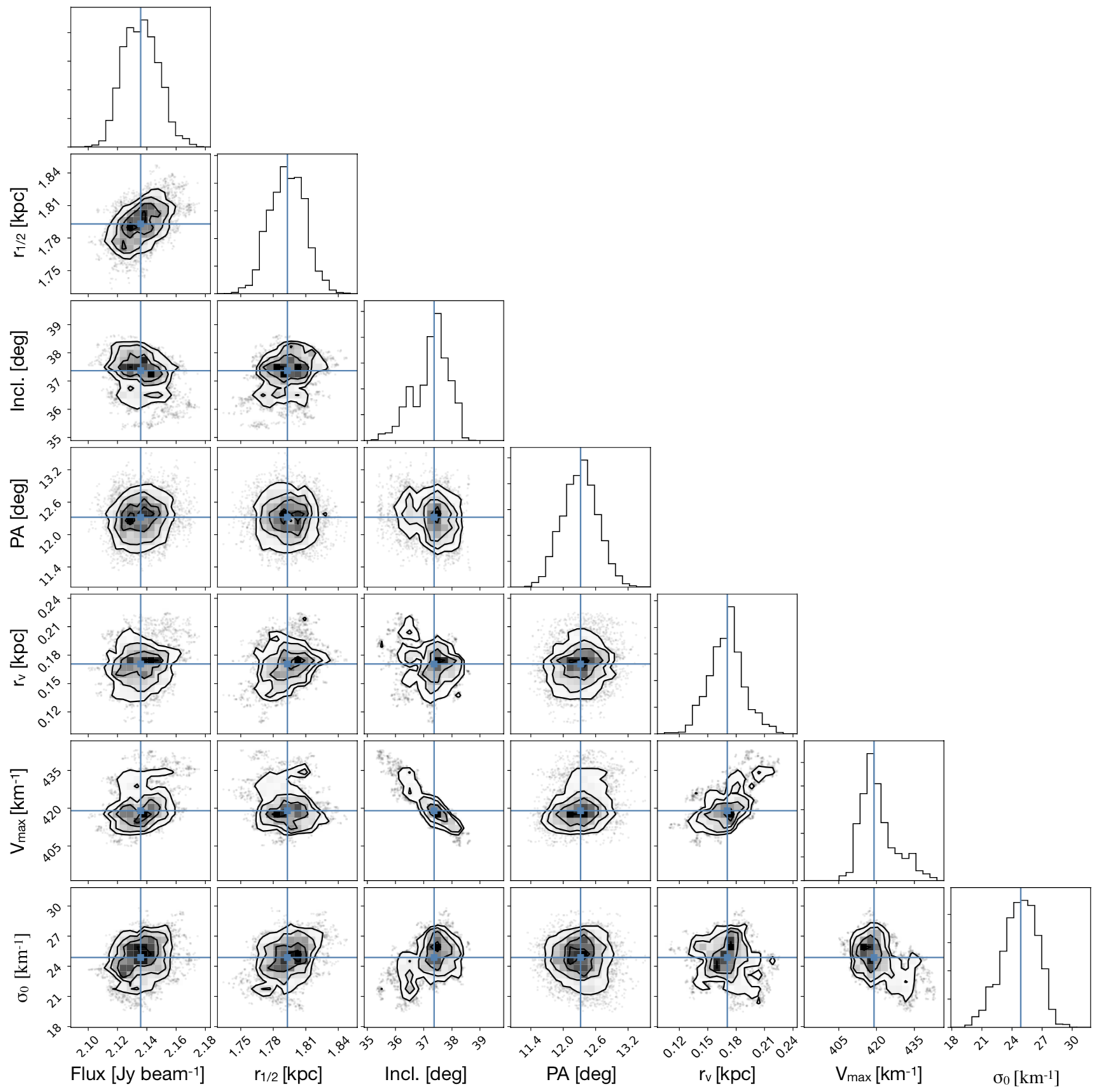}
        \caption{MCMC results for the SB1: the panels show the posterior probability distributions of seven model parameters using MCMC sampling with GalPAK$^{3D}$. Their marginalized probability distribution are shown as histograms. The parameters are the total flux (Flux), the disk half-light radius ($r_{\rm 1/2}$), the inclination angle (Incl.), the position angle (PA), the turnover radius ($r_{\rm v}$), the maximum circular velocity ($V_{\mathrm{max}}$), and the intrinsic velocity dispersion ($\sigma_{\mathrm{0}}$), in the same units as summarized in Table \ref{summary}. The solid blue lines show median values. The black contour corresponds to the 68\%, 95\%, and 99.7\% confidence intervals.}
        \label{MCMC_SB1}
\end{figure*}

\begin{figure*}[h!]
        \centering
        \includegraphics[scale=0.54]{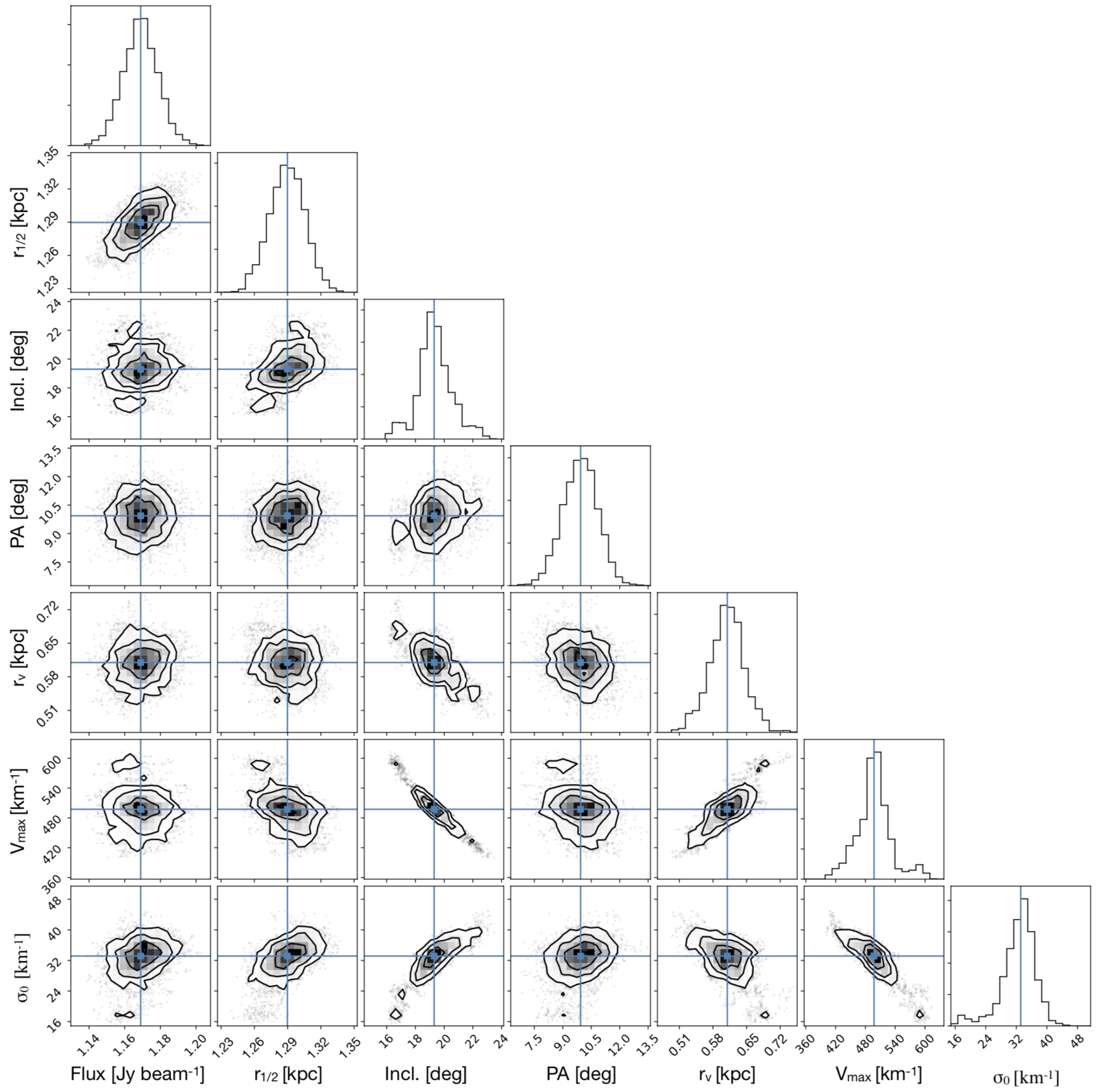}
        \caption{Same as Fig. \ref{MCMC_SB1}, but for the SB2.} 
        \label{MCMC_SB2}
\end{figure*}

\begin{figure*}[h!]
        \centering
        \includegraphics[scale=0.54]{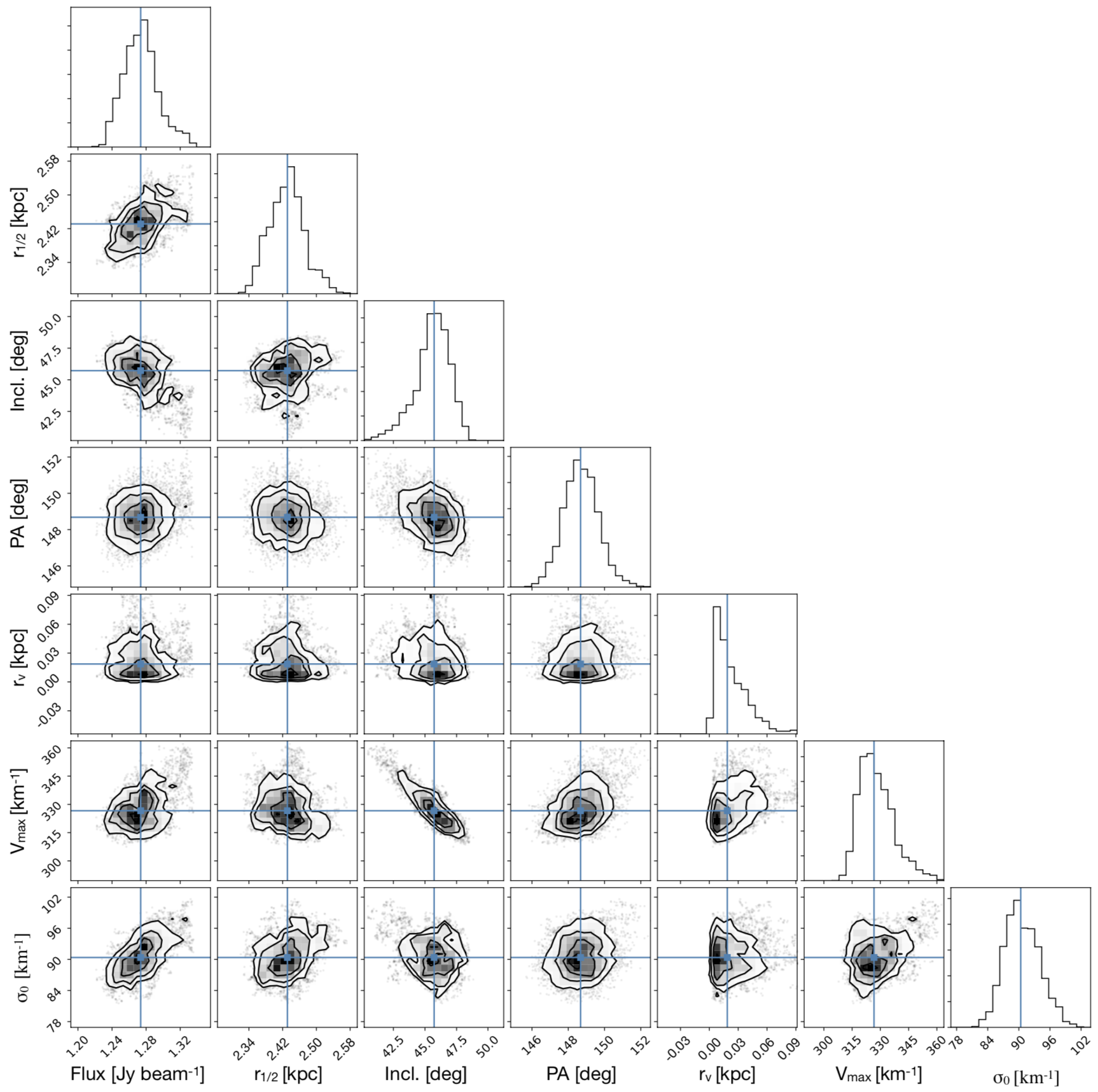}
        \caption{Same as Fig. \ref{MCMC_SB1}, but for the MS1.} 
        \label{MCMC_MS1}
\end{figure*}

\begin{figure*}[h!]
        \centering
        \includegraphics[scale=0.54]{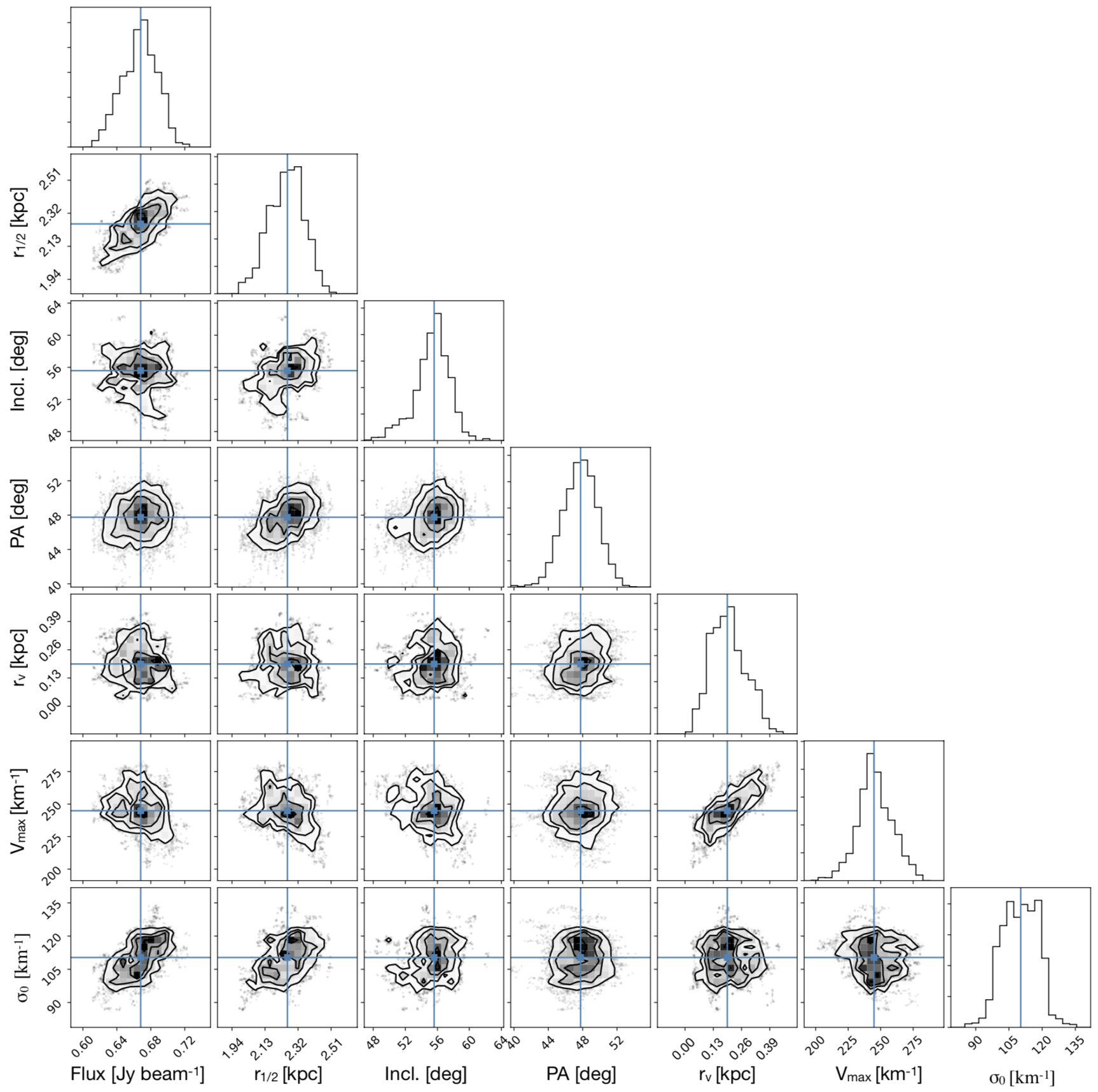}
        \caption{Same as Fig. \ref{MCMC_SB1}, but for the MS2.} 
        \label{MCMC_MS2}
\end{figure*}

\end{appendix}

\begin{thebibliography}{}
\bibitem[Aravena et al.(2014)]{Aravena2014} Aravena, M., Hodge, J.~A., Wagg, J., et al.\ 2014, \mnras, 442, 558. doi:10.1093/mnras/stu838
\bibitem[Barro et al.(2017)]{Barro2017} Barro, G., Kriek, M., P{\'e}rez-Gonz{\'a}lez, P.~G., et al.\ 2017, \apjl, 851, L40. doi:10.3847/2041-8213/aa9f0d
\bibitem[B{\'e}thermin et al.(2015)]{Bethermin2015} B{\'e}thermin, M., Daddi, E., Magdis, G., et al.\ 2015, \aap, 573, A113. doi:10.1051/0004-6361/201425031
\bibitem[Binney \& Tremaine(2008)]{Binney2008} Binney, J. \& Tremaine, S.\ 2008, Galactic Dynamics: Second Edition, by James Binney and Scott Tremaine. ISBN 978-0-691-13026-2 (HB). Published by Princeton University Press, Princeton, NJ USA, 2008.
\bibitem[Blain et al.(2004)]{Blain2004} Blain, A.~W., Chapman, S.~C., Smail, I., \& Ivison, R.\ 2004, \apj, 611, 725 
\bibitem[Boogaard et al.(2020)]{Boogaard2020} Boogaard, L.~A., van der Werf, P., Weiss, A., et al.\ 2020, \apj, 902, 109. doi:10.3847/1538-4357/abb82f
\bibitem[Boquien et al.(2019)]{Boquien2019} Boquien, M., Burgarella, D., Roehlly, Y., et al.\ 2019, \aap, 622, A103. doi:10.1051/0004-6361/201834156
\bibitem[Bothwell et al.(2013)]{Bothwell2013} Bothwell, M.~S., Smail, I., Chapman, S.~C., et al.\ 2013, \mnras, 429, 3047. doi:10.1093/mnras/sts562
\bibitem[Bouch{\'e} et al.(2015)]{Bouche2015} Bouch{\'e}, N., Carfantan, H., Schroetter, I., et al.\ 2015, \aj, 150, 92. doi:10.1088/0004-6256/150/3/92
\bibitem[Bournaud et al.(2011)]{Bournaud2011} Bournaud, F., Chapon, D., Teyssier, R., et al.\ 2011, \apj, 730, 4. doi:10.1088/0004-637X/730/1/4
\bibitem[Burgarella et al.(2005)]{Burgarella2005} Burgarella, D., Buat, V., \& Iglesias-P{\'a}ramo, J.\ 2005, \mnras, 360, 1413. doi:10.1111/j.1365-2966.2005.09131.x
\bibitem[Cacciato et al.(2012)]{Cacciato2012} Cacciato, M., Dekel, A., \& Genel, S.\ 2012, \mnras, 421, 818 
\bibitem[Calistro Rivera et al.(2018)]{Rivera2018} Calistro Rivera, G., Hodge, J.~A., Smail, I., et al.\ 2018, \apj, 863, 56. doi:10.3847/1538-4357/aacffa
\bibitem[Casey et al.(2015)]{Casey2015} Casey, C.~M., Cooray, A., Capak, P., et al.\ 2015, \apjl, 808, L33. doi:10.1088/2041-8205/808/2/L33
\bibitem[Casey(2016)]{Casey2016} Casey, C.~M.\ 2016, \apj, 824, 36. doi:10.3847/0004-637X/824/1/36
\bibitem[Chabrier(2003)]{Chabrier2003} Chabrier, G.\ 2003, \pasp, 115, 763. doi:10.1086/376392
\bibitem[Champagne et al.(2021)]{Champagne2021} Champagne, J.~B., Casey, C.~M., Zavala, J.~A., et al.\ 2021, \apj, 913, 110. doi:10.3847/1538-4357/abf4e6
\bibitem[Coogan et al.(2018)]{Coogan2018} Coogan, R.~T., Daddi, E., Sargent, M.~T., et al.\ 2018, \mnras, 479, 703
\bibitem[Cresci et al.(2009)]{Cresci2009} Cresci, G., Hicks, E.~K.~S., Genzel, R., et al.\ 2009, \apj, 697, 115. doi:10.1088/0004-637X/697/1/115
\bibitem[Cucciati et al.(2018)]{Cucciati2018} Cucciati, O., Lemaux, B.~C., Zamorani, G., et al.\ 2018, \aap, 619, A49. doi:10.1051/0004-6361/201833655
\bibitem[Daddi et al.(2015)]{Daddi2015} Daddi, E., Dannerbauer, H., Liu, D., et al.\ 2015, \aap, 577, A46. doi:10.1051/0004-6361/201425043
\bibitem[Daddi et al.(2017)]{Daddi2017} Daddi, E., Jin, S., Strazzullo, V., et al.\ 2017, \apjl, 846, L31. doi:10.3847/2041-8213/aa8808
\bibitem[Dannerbauer et al.(2014)]{Dannerbauer2014} Dannerbauer, H., Kurk, J.~D., De Breuck, C., et al.\ 2014, \aap, 570, A55 
\bibitem[Dannerbauer et al.(2017)]{Dannerbauer2017} Dannerbauer, H., Lehnert, M.~D., Emonts, B., et al.\ 2017, \aap, 608, A48. doi:10.1051/0004-6361/201730449
\bibitem[Danovich et al.(2015)]{Danovich2015} Danovich, M., Dekel, A., Hahn, O., et al.\ 2015, \mnras, 449, 2087. doi:10.1093/mnras/stv270
\bibitem[Danovich et al.(2012)]{Danovich2012} Danovich, M., Dekel, A., Hahn, O., et al.\ 2012, \mnras, 422, 1732. doi:10.1111/j.1365-2966.2012.20751.x
\bibitem[Dekel et al.(2020)]{Dekel2020a} Dekel, A., Ginzburg, O., Jiang, F., et al.\ 2020, \mnras, 493, 4126. doi:10.1093/mnras/staa470
\bibitem[Di Teodoro et al.(2016)]{Di Teodoro2016} Di Teodoro, E.~M., Fraternali, F., \& Miller, S.~H.\ 2016, \aap, 594, A77. doi:10.1051/0004-6361/201628315
\bibitem[Di Teodoro \& Fraternali(2015)]{Di Teodoro2015} Di Teodoro, E.~M. \& Fraternali, F.\ 2015, \mnras, 451, 3021. doi:10.1093/mnras/stv1213
\bibitem[Downes \& Solomon(1998)]{Downes1998} Downes, D. \& Solomon, P.~M.\ 1998, \apj, 507, 615. doi:10.1086/306339
\bibitem[Draine et al.(2014)]{Draine2014} Draine, B.~T., Aniano, G., Krause, O., et al.\ 2014, \apj, 780, 172. doi:10.1088/0004-637X/780/2/172
\bibitem[Dressler(1980)]{Dressler1980} Dressler, A.\ 1980, \apj, 236, 351. doi:10.1086/157753
\bibitem[Elbaz et al.(2018)]{Elbaz2018} Elbaz, D., Leiton, R., Nagar, N., et al.\ 2018, \aap, 616, A110. doi:10.1051/0004-6361/201732370
\bibitem[Epinat et al.(2012)]{Epinat2012} Epinat, B., Tasca, L., Amram, P., et al.\ 2012, \aap, 539, A92. doi:10.1051/0004-6361/201117711
\bibitem[Epinat et al.(2010)]{Epinat2010} Epinat, B., Amram, P., Balkowski, C., et al.\ 2010, \mnras, 401, 2113. doi:10.1111/j.1365-2966.2009.15688.x
\bibitem[Fensch et al.(2017)]{Fensch2017} Fensch, J., Renaud, F., Bournaud, F., et al.\ 2017, \mnras, 465, 1934 
\bibitem[Fraternali et al.(2021)]{Fraternali2021} Fraternali, F., Karim, A., Magnelli, B., et al.\ 2021, \aap, 647, A194. doi:10.1051/0004-6361/202039807
\bibitem[Fritz et al.(2006)]{Fritz2006} Fritz, J., Franceschini, A., \& Hatziminaoglou, E.\ 2006, \mnras, 366, 767. doi:10.1111/j.1365-2966.2006.09866.x
\bibitem[F{\"o}rster Schreiber et al.(2009)]{Forster2009} F{\"o}rster Schreiber, N.~M., Genzel, R., Bouch{\'e}, N., et al.\ 2009, \apj, 706, 1364. doi:10.1088/0004-637X/706/2/1364
\bibitem[F{\"o}rster Schreiber \& Wuyts(2020)]{Forster-Schreiber2020} F{\"o}rster Schreiber, N.~M. \& Wuyts, S.\ 2020, \araa, 58, 661. doi:10.1146/annurev-astro-032620-021910
\bibitem[Fujimoto et al.(2021)]{Fujimoto2021} Fujimoto, S., Oguri, M., Brammer, G., et al.\ 2021, \apj, 911, 99. doi:10.3847/1538-4357/abd7ec
\bibitem[Genzel et al.(2015)]{Genzel2015} Genzel, R., Tacconi, L.~J., Lutz, D., et al.\ 2015, \apj, 800, 20. doi:10.1088/0004-637X/800/1/20
\bibitem[Gnerucci et al.(2011)]{Gnerucci2011} Gnerucci, A., Marconi, A., Cresci, G., et al.\ 2011, \aap, 528, A88. doi:10.1051/0004-6361/201015465
\bibitem[Green et al.(2014)]{Green2014} Green, A.~W., Glazebrook, K., McGregor, P.~J., et al.\ 2014, \mnras, 437, 1070. doi:10.1093/mnras/stt1882
\bibitem[G{\'o}mez-Guijarro et al.(2019)]{Gomez-Guijarro2019} G{\'o}mez-Guijarro, C., Riechers, D.~A., Pavesi, R., et al.\ 2019, \apj, 872, 117. doi:10.3847/1538-4357/ab002a
\bibitem[G{\'o}mez-Guijarro et al.(2022a)]{Gomez-Guijarro2021} G{\'o}mez-Guijarro, C., Elbaz, D., Xiao, M., et al.\ 2022a, \aap, 658, A43. doi:10.1051/0004-6361/202141615
\bibitem[G{\'o}mez-Guijarro et al.(2022b)]{Carlos2022} G{\'o}mez-Guijarro, C., Elbaz, D., Xiao, M., et al.\ 2022b, \aap, 659, A196. doi:10.1051/0004-6361/202142352
\bibitem[Habing(1968)]{Habing1968} Habing, H.~J.\ 1968, \bain, 19, 421
\bibitem[Harrington et al.(2018)]{Harrington2018} Harrington, K.~C., Yun, M.~S., Magnelli, B., et al.\ 2018, \mnras, 474, 3866. doi:10.1093/mnras/stx3043
\bibitem[Hayashi et al.(2016)]{Hayashi2016} Hayashi, M., Kodama, T., Tanaka, I., et al.\ 2016, \apjl, 826, L28  
\bibitem[Hodge et al.(2012)]{Hodge2012} Hodge, J.~A., Carilli, C.~L., Walter, F., et al.\ 2012, \apj, 760, 11 
\bibitem[Hodge et al.(2013)]{Hodge2013} Hodge, J.~A., Carilli, C.~L., Walter, F., et al.\ 2013, \apj, 776, 22
\bibitem[Hopkins et al.(2009)]{Hopkins2009} Hopkins, P.~F., Cox, T.~J., Younger, J.~D., \& Hernquist, L.\ 2009, \apj, 691, 1168 
\bibitem[Johnson et al.(2018)]{Johnson2018} Johnson, H.~L., Harrison, C.~M., Swinbank, A.~M., et al.\ 2018, \mnras, 474, 5076. doi:10.1093/mnras/stx3016
\bibitem[Kennicutt \& Evans(2012)]{kennicutt} Kennicutt, R.~C. \& Evans, N.~J.\ 2012, \araa, 50, 531. doi:10.1146/annurev-astro-081811-125610
\bibitem[Kohandel et al.(2020)]{Kohandel2020} Kohandel, M., Pallottini, A., Ferrara, A., et al.\ 2020, \mnras, 499, 1250. doi:10.1093/mnras/staa2792
\bibitem[Kravtsov \& Borgani(2012)]{Kravtsov2012} Kravtsov, A.~V. \& Borgani, S.\ 2012, \araa, 50, 353. doi:10.1146/annurev-astro-081811-125502
\bibitem[Kretschmer et al.(2020)]{Kretschmer2020} Kretschmer, M., Agertz, O., \& Teyssier, R.\ 2020, \mnras, 497, 4346. doi:10.1093/mnras/staa2243
\bibitem[Kretschmer et al.(2021)]{Kretschmer2021} Kretschmer, M., Dekel, A., Freundlich, J., et al.\ 2021, \mnras, 503, 5238. doi:10.1093/mnras/stab833
\bibitem[Krumholz et al.(2018)]{Krumholz2018} Krumholz, M.~R., Burkhart, B., Forbes, J.~C., et al.\ 2018, \mnras, 477, 2716. doi:10.1093/mnras/sty852
\bibitem[Krumholz \& Burkhart(2016)]{Krumholz2016} Krumholz, M.~R. \& Burkhart, B.\ 2016, \mnras, 458, 1671. doi:10.1093/mnras/stw434
\bibitem[Kroupa(2001)]{Kroupa2001} Kroupa, P.\ 2001, \mnras, 322, 231. doi:10.1046/j.1365-8711.2001.04022.x
\bibitem[Law et al.(2009)]{Law2009} Law, D.~R., Steidel, C.~C., Erb, D.~K., et al.\ 2009, \apj, 697, 2057. doi:10.1088/0004-637X/697/2/2057
\bibitem[Lelli et al.(2021)]{Lelli2021} Lelli, F., Di Teodoro, E.~M., Fraternali, F., et al.\ 2021, Science, 371, 713. doi:10.1126/science.abc1893
\bibitem[Leroy et al.(2008)]{Leroy2008} Leroy, A.~K., Walter, F., Brinks, E., et al.\ 2008, \aj, 136, 2782. doi:10.1088/0004-6256/136/6/2782
\bibitem[Leroy et al.(2009)]{Leroy2009} Leroy, A.~K., Walter, F., Bigiel, F., et al.\ 2009, \aj, 137, 4670. doi:10.1088/0004-6256/137/6/4670
\bibitem[Leroy et al.(2011)]{Leroy2011} Leroy, A.~K., Bolatto, A., Gordon, K., et al.\ 2011, \apj, 737, 12. doi:10.1088/0004-637X/737/1/12
\bibitem[Li et al.(2021)]{Li2021} Li, Q., Wang, R., Dannerbauer, H., et al.\ 2021, \apj, 922, 236. doi:10.3847/1538-4357/ac29c6
\bibitem[Liu et al.(2019)]{Liu2019} Liu, D., Schinnerer, E., Groves, B., et al.\ 2019, \apj, 887, 235. doi:10.3847/1538-4357/ab578d
\bibitem[Madau \& Dickinson(2014)]{Madau2014} Madau, P. \& Dickinson, M.\ 2014, \araa, 52, 415. doi:10.1146/annurev-astro-081811-125615
\bibitem[Magdis et al.(2011)]{Magdis2011} Magdis, G.~E., Daddi, E., Elbaz, D., et al.\ 2011, \apjl, 740, L15. doi:10.1088/2041-8205/740/1/L15
\bibitem[Magdis et al.(2012)]{Magdis2012} Magdis, G.~E., Daddi, E., B{\'e}thermin, M., et al.\ 2012, \apj, 760, 6. doi:10.1088/0004-637X/760/1/6
\bibitem[McMullin et al.(2007)]{mcmullin} McMullin, J.~P., Waters, B., Schiebel, D., et al.\ 2007, Astronomical Data Analysis Software and Systems XVI, 376, 127
\bibitem[Miller et al.(2018)]{Miller2018} Miller, T.~B., Chapman, S.~C., Aravena, M., et al.\ 2018, \nat, 556, 469. doi:10.1038/s41586-018-0025-2
\bibitem[Muzzin et al.(2013)]{Muzzin2013} Muzzin, A., Marchesini, D., Stefanon, M., et al.\ 2013, \apjs, 206, 8. doi:10.1088/0067-0049/206/1/8
\bibitem[Noll et al.(2009)]{Noll2009} Noll, S., Burgarella, D., Giovannoli, E., et al.\ 2009, \aap, 507, 1793. doi:10.1051/0004-6361/200912497
\bibitem[Oteo et al.(2018)]{Oteo2018} Oteo, I., Ivison, R.~J., Dunne, L., et al.\ 2018, \apj, 856, 72. doi:10.3847/1538-4357/aaa1f1
\bibitem[Peng et al.(2010)]{Peng2010} Peng, Y.-. jie ., Lilly, S.~J., Kova{\v{c}}, K., et al.\ 2010, \apj, 721, 193. doi:10.1088/0004-637X/721/1/193
\bibitem[Perna et al.(2022)]{Perna2022} Perna, M., Arribas, S., Colina, L., et al.\ 2022, arXiv:2202.02336
\bibitem[Pettini \& Pagel(2004)]{Pettini2004} Pettini, M. \& Pagel, B.~E.~J.\ 2004, \mnras, 348, L59. doi:10.1111/j.1365-2966.2004.07591.x
\bibitem[Pillepich et al.(2019)]{pillepich} Pillepich, A., Nelson, D., Springel, V., et al.\ 2019, \mnras, 490, 3196. doi:10.1093/mnras/stz2338
\bibitem[Puglisi et al.(2019)]{Puglisi2019} Puglisi, A., Daddi, E., Liu, D., et al.\ 2019, \apjl, 877, L23. doi:10.3847/2041-8213/ab1f92
\bibitem[Puglisi et al.(2021)]{Puglisi2021} Puglisi, A., Daddi, E., Valentino, F., et al.\ 2021, \mnras, 508, 5217. doi:10.1093/mnras/stab2914
\bibitem[Riechers et al.(2010)]{Riechers2010} Riechers, D.~A., Carilli, C.~L., Walter, F., et al.\ 2010, \apjl, 724, L153. doi:10.1088/2041-8205/724/2/L153
\bibitem[Riechers et al.(2020)]{Riechers2020} Riechers, D.~A., Boogaard, L.~A., Decarli, R., et al.\ 2020, \apjl, 896, L21. doi:10.3847/2041-8213/ab9595
\bibitem[Rizzo et al.(2020)]{Rizzo2020} Rizzo, F., Vegetti, S., Powell, D., et al.\ 2020, \nat, 584, 201. doi:10.1038/s41586-020-2572-6
\bibitem[Rizzo et al.(2021)]{Rizzo2021} Rizzo, F., Vegetti, S., Fraternali, F., et al.\ 2021, \mnras, 507, 3952. doi:10.1093/mnras/stab2295
\bibitem[Robertson et al.(2006)]{Robertson2006} Robertson, B., Bullock, J.~S., Cox, T.~J., et al.\ 2006, \apj, 645, 986 
\bibitem[R{\'e}my-Ruyer et al.(2014)]{Remy-Ruyer2014} R{\'e}my-Ruyer, A., Madden, S.~C., Galliano, F., et al.\ 2014, \aap, 563, A31. doi:10.1051/0004-6361/201322803
\bibitem[Salpeter(1955)]{Salpeter1955} Salpeter, E.~E.\ 1955, \apj, 121, 161. doi:10.1086/145971
\bibitem[Schreiber et al.(2015)]{Schreiber2015} Schreiber, C., Pannella, M., Elbaz, D., et al.\ 2015, \aap, 575, A74. doi:10.1051/0004-6361/201425017
\bibitem[Scoville et al.(2017)]{Scoville2017} Scoville, N., Lee, N., Vanden Bout, P., et al.\ 2017, \apj, 837, 150. doi:10.3847/1538-4357/aa61a0
\bibitem[Scoville et al.(2016)]{Scoville2016} Scoville, N., Sheth, K., Aussel, H., et al.\ 2016, \apj, 820, 83. doi:10.3847/0004-637X/820/2/83
\bibitem[Sharon et al.(2013)]{Sharon2013} Sharon, C.~E., Baker, A.~J., Harris, A.~I., et al.\ 2013, \apj, 765, 6. doi:10.1088/0004-637X/765/1/6
\bibitem[Sharon et al.(2016)]{Sharon2016} Sharon, C.~E., Riechers, D.~A., Hodge, J., et al.\ 2016, \apj, 827, 18. doi:10.3847/0004-637X/827/1/18
\bibitem[Simons et al.(2016)]{Simons2016} Simons, R.~C., Kassin, S.~A., Trump, J.~R., et al.\ 2016, \apj, 830, 14. doi:10.3847/0004-637X/830/1/14
\bibitem[Solomon et al.(1997)]{Solomon1997} Solomon, P.~M., Downes, D., Radford, S.~J.~E., et al.\ 1997, \apj, 478, 144. doi:10.1086/303765
\bibitem[Solomon \& Vanden Bout(2005)]{Solomon2005} Solomon, P.~M. \& Vanden Bout, P.~A.\ 2005, \araa, 43, 677. doi:10.1146/annurev.astro.43.051804.102221
\bibitem[Swinbank et al.(2017)]{Swinbank2017} Swinbank, A.~M., Harrison, C.~M., Trayford, J., et al.\ 2017, \mnras, 467, 3140. doi:10.1093/mnras/stx201
\bibitem[Swinbank et al.(2011)]{Swinbank2011} Swinbank, A.~M., Papadopoulos, P.~P., Cox, P., et al.\ 2011, \apj, 742, 11. doi:10.1088/0004-637X/742/1/11
\bibitem[Tacconi et al.(2018)]{Tacconi2018} Tacconi, L.~J., Genzel, R., Saintonge, A., et al.\ 2018, \apj, 853, 179. doi:10.3847/1538-4357/aaa4b4
\bibitem[Tacconi et al.(2013)]{Tacconi2013} Tacconi, L.~J., Neri, R., Genzel, R., et al.\ 2013, \apj, 768, 74. doi:10.1088/0004-637X/768/1/74
\bibitem[Tacconi et al.(2008)]{Tacconi2008} Tacconi, L.~J., Genzel, R., Smail, I., et al.\ 2008, \apj, 680, 246. doi:10.1086/587168
\bibitem[Tadaki et al.(2017)]{Tadaki2017} Tadaki, K.-. ichi ., Kodama, T., Nelson, E.~J., et al.\ 2017, \apjl, 841, L25. doi:10.3847/2041-8213/aa7338
\bibitem[Tadaki et al.(2019)]{Tadaki2019} Tadaki, K.-. ichi ., Iono, D., Hatsukade, B., et al.\ 2019, \apj, 876, 1. doi:10.3847/1538-4357/ab1415
\bibitem[Thompson et al.(2005)]{Thompson2005} Thompson, T.~A., Quataert, E., \& Murray, N.\ 2005, \apj, 630, 167 
\bibitem[Toomre(1964)]{toomre} Toomre, A.\ 1964, \apj, 139, 1217. doi:10.1086/147861
\bibitem[Turner et al.(2017)]{Turner2017} Turner, O.~J., Cirasuolo, M., Harrison, C.~M., et al.\ 2017, \mnras, 471, 1280. doi:10.1093/mnras/stx1366
\bibitem[Ueda et al.(2014)]{Ueda2014} Ueda, J., Iono, D., Yun, M.~S., et al.\ 2014, \apjs, 214, 1 
\bibitem[Valentino et al.(2020)]{Valentino2020} Valentino, F., Daddi, E., Puglisi, A., et al.\ 2020, \aap, 641, A155. doi:10.1051/0004-6361/202038322
\bibitem[van der Wel et al.(2014)]{vanderWel2014} van der Wel, A., Franx, M., van Dokkum, P.~G., et al.\ 2014, \apj, 788, 28. doi:10.1088/0004-637X/788/1/28
\bibitem[Wang et al.(2016)]{Wang2016} Wang, T., Elbaz, D., Daddi, E., et al.\ 2016, \apj, 828, 56. doi:10.3847/0004-637X/828/1/56
\bibitem[Wang et al.(2018)]{Wang2018} Wang, T., Elbaz, D., Daddi, E., et al.\ 2018, \apjl, 867, L29. doi:10.3847/2041-8213/aaeb2c
\bibitem[Wisnioski et al.(2015)]{Wisnioski2015} Wisnioski, E., F{\"o}rster Schreiber, N.~M., Wuyts, S., et al.\ 2015, \apj, 799, 209. doi:10.1088/0004-637X/799/2/209
\bibitem[Wisnioski et al.(2019)]{Wisnioski2019} Wisnioski, E., F{\"o}rster Schreiber, N.~M., Fossati, M., et al.\ 2019, \apj, 886, 124. doi:10.3847/1538-4357/ab4db8
\bibitem[{\"U}bler et al.(2019)]{Ubler2019} {\"U}bler, H., Genzel, R., Wisnioski, E., et al.\ 2019, \apj, 880, 48. doi:10.3847/1538-4357/ab27cc


\end{thebibliography}
\end{document}